\documentclass[prb,superscriptaddress,twocolumn,
 amsmath,amssymb,
 aps,
]{revtex4-2}

\usepackage{graphicx}
\usepackage{dcolumn}
\usepackage{bm}
\usepackage{hyperref}
\usepackage{xcolor}
\usepackage{physics}
\usepackage{float}
\usepackage{overpic}

\newcommand{\mr}[1]{\mathrm{#1}}
\newcommand{\dg}{\dagger}

\begin{document}

\title{Two-dimensional many-body localized systems coupled to a heat bath}

\author{Joey Li}
\affiliation{Institute for Theoretical Physics, University of Innsbruck, 6020 Innsbruck, Austria}
\affiliation{Institute for Quantum Optics and Quantum Information of the Austrian Academy of Sciences, 6020 Innsbruck, Austria}

\author{Amos Chan}
\affiliation{Department of Physics, Lancaster University, Lancaster LA1 4YB, United Kingdom}

\author{Thorsten B. Wahl}
\affiliation{TCM Group, Cavendish Laboratory, Department of Physics, J J Thomson Avenue, Cambridge CB3 0HE, United Kingdom}

\date{\today}

\begin{abstract}
We numerically investigate the effect of coupling a two-dimensional many-body localized system to a finite heat bath, using shallow quantum circuits as a variational ansatz. Specifically, we simulate optical lattice experiments with two components of ultracold bosons, where only one species is subject to a random disorder potential and the other acts as a heat bath. We obtain a filling fraction-dependent phase diagram with a critical filling consistent with experiments. We also perform long-time time-dependent variational principle (TDVP) simulations with matrix product states, but find that the quantum circuit approach produces superior results due to the lack of boundary effects. Finally, we present additional results on the two-point correlation functions and the quantum mutual information between sites.
\end{abstract}
\maketitle

\section{Introduction}

The assumption that isolated systems act as their own heat baths and therefore reach thermal equilibrium was largely taken for granted until the beginning of this century when several analytical~\cite{gornyi2005interacting,basko2006metal} and numerical works~\cite{znidaric2008many,pal2010mb} started to provide evidence for the existence of many-body localization (MBL). Experiments observing MBL~\cite{Schreiber842,Smith_MBL,Choi2017,Roushan2017,Lukin2018} and a rigorous proof~\cite{imbrie2016many} in the last decade seemed to put the phenomenon on solid ground. However, in recent years, multiple works started questioning the existence of true MBL~\cite{untajs2020,Sels2021,Peacock2023}, even in one dimension. This would imply that MBL in one dimension is only a transient effect (as has long been conjectured for higher dimensions~\cite{deRoeck2017Stability,Gopalakrishnan2019}), though on incredibly long time scales.

Extremely long thermalization times imply that strongly disordered isolated quantum systems behave many-body localized on experimentally relevant time scales. That is, for all practical purposes one can assume the existence of MBL, even in higher dimensions, where it has also been experimentally observed~\cite{Choi1547}. In this spirit, we investigate the effect of coupling a localized and delocalized system in higher dimensions, a subject that is relevant for the understanding of thermalization phenomena and time scales. In particular, we study a two-dimensional ``MBL-like" system coupled to a heat bath, and aim to reproduce the phase transition point found in two-dimensional optical lattice experiments with two-component ultracold bosons~\cite{cleanDirty2DMBL} as the size of the heat bath is increased. In those experiments, one boson species was subject to random disorder while the other only experienced the optical lattice potential, acting as a heat bath for the former. The experiments were also theoretically analyzed in Ref.~\cite{prabhu2021bathmediateddecaydensity}, albeit using a product state ansatz that assumes the observations were caused by glassy dynamics~\cite{glassyMBL}. 

We use quantum circuits, a type of tensor networks, as a variational ansatz for our classical simulations. Shallow quantum circuits have proven useful for both the numerical description of one-~\cite{Pollmann2016TNS,Wahl2017PRX,Wahl2022} and two-dimensional MBL systems~\cite{2DMBL,2dMBLfermion,Venn2023} and the rigorous classification of symmetry-protected topological MBL phases~\cite{Thorsten,1DSPTMBL,2DSPTMBL}. We optimize a two-dimensional shallow quantum circuit to approximately diagonalize a hardcore version of the Hamiltonian modeling the experiments of Ref.~\cite{cleanDirty2DMBL}. The unitaries building up the quantum circuit contain the variational parameters of the approach. After optimizing those, we have an approximate representation of the overall unitary diagonalizing the Hamiltonian and can extract local features (such as correlations) of its eigenstates. We analyze entanglement features in the different particle sectors of the diagonalizing unitary to obtain the phase diagram 
as a function of the disorder strength and the fraction of bosons not experiencing the disorder potential. [Note that we use the semantics of a phase transition throughout this work, although transience of MBL implies that instead we have a (relatively sharp) crossover between a regime of extremely slow, experimentally inaccessible thermalization and fast thermalization.] This phase diagram constitutes the central novel finding of our work. We obtain a phase transition point in the same regime as the measured value. 

\begin{figure*}[t!]
\centering
\includegraphics[width=0.96 \textwidth  ]{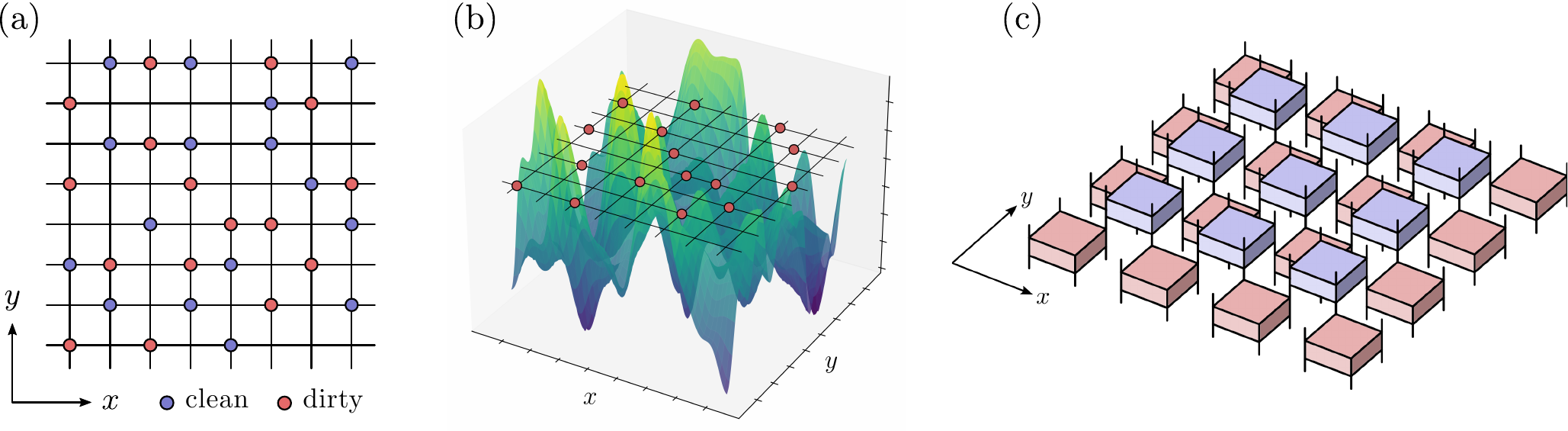} 
\caption{(a) Optical lattice filled with two species of bosons: ``clean" (blue) and ``dirty" (red).  Half the lattice sites are occupied: the overall filling is $\nu=0.5$. We consider various relative fractions $f_c$ and $f_d$ of clean and dirty particles: depicted here is the case $f_c = f_d = 0.5$.  (b) The system is in a random potential that affects \textit{only} the dirty particles. (c) Two-layer quantum circuit $U$ used as a variational ansatz to approximately diagonalize the Hamiltonian.  (In this work we consider system sizes $L=6$ and $L=10$.) }
     \label{fig:combined}
\end{figure*}

For comparison, we also simulated real time evolution with matrix product states using the time-dependent variational principle~\cite{tdvp,Elmer}. 
We found that the requirement of numerically accurate time evolutions restricts us to fairly small system sizes ($6 \times 6$) with open boundary conditions. As a result, boundary effects and the shift of the two-dimensional MBL-to-thermal transition point with system size~\cite{Gopalakrishnan2019,Elmer} lead to significantly lower critical disorder strengths. However, the corresponding phase diagram is qualitatively the same as the one obtained with quantum circuits, which do not suffer from these restrictions (but rather finite-entanglement effects). We thus find that our method is more suitable to describe the limit of large, experimentally relevant, system sizes. 

Furthermore, we extracted the correlation lengths from our optimized unitaries, which show a slow decay with disorder strength. Finally, we find that the quantum mutual information distributions between sites separated by short and medium distances 
develop longer tails as the disorder strength is lowered towards the transition point. If similar features are found in future, more accurate studies, then resonances are also at play on experimentally relevant time scales near the transition in higher dimensions, similar to one dimensional MBL~\cite{Crowley2022}.

\section{Model and numerical approach}

We study a two-dimensional MBL system coupled to an immersed heat bath. Specifically, we consider two components of bosons confined by an optical square lattice. One species of bosons (``dirty bosons'') experiences a randomly chosen on-site disorder potential while the other one (``clean bosons'') acts as a heat bath. We assume that the dynamics of the bosons is governed by the Hamiltonian of Ref.~\cite{cleanDirty2DMBL}, 
\begin{align}
    H_\mr{exp} = &-J \sum_{\langle i,j \rangle, \sigma} {a}^\dagger_{i,\sigma} {a}_{j,\sigma} + \frac{V}{2} \sum_{i,\sigma} {n}_{i,\sigma} ({n}_{i,\sigma} - 1) \nonumber \\
    &+ V \sum_i {n}_{i,d} {n}_{i,c} + \sum_i \delta_i {n}_{i,d},
\end{align}
where $i = (x,y)$ denotes the lattice site. $J$ is the tunneling amplitude, i.e., hopping strength, $V$ the on-site repulsion strength, and $\delta_{i}$ the random disorder potential, which is chosen according to a Gaussian distribution with half-maximum width $\Delta$. We consider only bosons in the center of the optical trap, where the trap potential can be neglected. $a_{i, \sigma}$ denotes the annihilation operator of a boson at site $i$ for $\sigma \in\{c,d\}$ with $c$ ($d$) corresponding to clean (dirty) bosons and $n_{i, \sigma} \equiv {a_{i, \sigma}^\dagger} a_{i, \sigma}$ the corresponding particle number operator. See Fig.~\ref{fig:combined} for an illustration. We define $f_c$ as the relative fraction of clean bosons, and $f_d$ as that of dirty bosons, where $f_c + f_d =1$.

In numerical simulations, one typically has to truncate the on-site occupation number of bosons, which can in principle be arbitrarily large. As experimentally only doublon fractions of up to $18 \, \%$ are observed~\cite{cleanDirty2DMBL} (due to the large on-site repulsion of $V=24.4J$) and to make our simulations numerically feasible, we restrict to the hardcore boson case. Thus, the model we study is
\begin{align}
    H = -J\sum_{\langle i,j \rangle, \sigma} {a}^\dagger_{i,\sigma} {a}_{j,\sigma} + \sum_i \delta_i {n}_{i,d}. \label{eq:Ham}
\end{align}
We note that due to the (infinite) on-site repulsion, this Hamiltonian still describes an interacting model. We work with an $L\times L=10\times10$ system with periodic boundary conditions. (We also consider, for comparison, the case $L=6$).

\subsection{Numerical methods}
Our study involves variationally diagonalizing the Hamiltonian~\eqref{eq:Ham}, i.e. obtaining an approximation of the full set of eigenstates. (Independently, we also simulate time evolution using TDVP; this forms the basis for the results in Section \ref{sec:tdvp}.) 

We use an ansatz $U$ that is a two-layer tensor network, or quantum circuit~\cite{2DMBL,Venn2023,2dMBLfermion}, with the structure as depicted in Fig.~\ref{fig:combined}(c). The ansatz is essentially a two-dimensional matrix product operator (MPO), or a generalization of projected entangled pair states (PEPS) that contains all eigenstates. 

We numerically optimize $U$ such that $UHU^\dagger$ is approximately diagonal. To that end, we minimize a cost function which is a sum of local tensor network contractions.  Since the Hamiltonian~\eqref{eq:Ham} preserves the particle numbers of both clean and dirty bosons, its eigenstates also have fixed clean and dirty particle numbers. We impose this at the level of the individual unitaries building up the quantum circuit in our numerical approach~\cite{2DMBL,2dMBLfermion}. This makes it straightforward to access individual particle sectors of our overall unitary.  For further details about the ansatz, see Appendix~\ref{app:sec1}. 

\subsubsection{Cost function}
We perform the variational diagonalization by minimizing a cost function, which can be motivated by the concept of local integrals of motion (LIOMs)~\cite{serbyn2013local,Huse_MBL_phenom_14,chandran2015constructing,ros2015integrals,Inglis_PRL2016,Rademaker2016LIOM,Monthus2016,Goihl2018,Abi2017}, also known as l-bits.  MBL systems are known to possess a set of LIOMs $\{\tau_i\}$, that (in two dimensions approximately~\cite{chandran2016higherD}) commute with the Hamiltonian and each other, i.e. $[H,\tau_i] = [\tau_i,\tau_j]=0$.  Each $\tau_i$ is an operator that is localized around site $i$, and is defined by $\tau_i = U_\mr{exact} \sigma^z_i U_\mr{exact}^\dagger$ for the case of spin-$1/2$ spin models such as the ``standard" disordered Heisenberg model~\cite{znidaric2008many,pal2010mb,Luitz2015}, or analogously otherwise. $U_\mr{exact}$ is the local unitary operator that diagonalizes the Hamiltonian.  Here we define a set of \textit{approximate} LIOMs given by $\tilde \tau_{i,\mu}(U) = Un_{i,\mu}U^\dagger$, where $\mu=c,d$, and $U$ is the variational quantum circuit.  To optimize $U$, we define a cost function $f\propto\sum_{i,\mu} \|[H, \tilde \tau_{i,\mu}]\|^2$ that penalizes the non-commutation of $H$ and $\tilde \tau_{i,\mu}$. (The $\tilde \tau_{i,\mu}$ already commute with each other by construction.) Thus, by minimizing the cost function we are optimizing the approximate LIOMs. We note that although the motivation was physical, it is also possible to mathematically justify this choice of cost function by showing from the definition of $\tilde \tau_{i,\mu}$ that $[H,\tilde \tau_{i,\mu}]=0$ for all $\tilde \tau_{i,\mu}$ if and only if $UHU^\dagger$ is diagonal. 

Specifically, the cost function $f(U)$ is given by
\begin{align}\label{eq:cost_fn}
    f(U) &= \frac{1}{2L^23^{L^2}}\sum_i \sum_{\mu = c,d} \Tr([H, \tilde \tau_{i,\mu}]^\dagger[H, \tilde \tau_{i,\mu}]) \notag\\
    &= \frac{1}{L^23^{L^2}}\sum_i \sum_{\mu = c,d} \big[ \Tr(H^2(\tilde \tau_{\mu,i})^2) - \Tr((H \tilde \tau_{i,\mu})^2)\big], 
\end{align}
where the factor of $1/2L^23^{L^2}$ has been included so that $f(U)$ is an $\mathcal{O}(1)$ number. Importantly, $f(U)$ is a sum of quantities that only involve local causal cone calculations, so the total computational cost scales linearly with the system size. We illustrate the tensor network contraction needed to calculate the cost function in Fig.~\ref{fig:rhoCalculations}(a) in the appendix. For further details on the calculation, we refer the reader to Refs.~\cite{2DMBL,2dMBLfermion}, where a similar optimization procedure was employed. 

\subsection{Approximate eigenstates}

We note that the approximate eigenstates encoded in our unitary $U$ represent the dynamics of the Hamiltonian $H$ on short time scales: it does not make a significant difference whether we consider the exact evolution of an initial state $|\psi_0\rangle$ via $|\psi(t) \rangle = e^{iHt}|\psi_0\rangle = U_\mr{exact} e^{i E_\mr{exact}t} U_\mr{exact}^\dg |\psi_0\rangle$ or the approximate evolution $|\tilde \psi(t)\rangle = U e^{i Et}U^\dg |\psi_0\rangle$, as long as the time $t$ is small, that is, less than $1/\| [H, \tilde \tau_i^z\|_\mathrm{op}$. (Here $E_\mr{exact} = U_\mr{exact}^\dg H U_\mr{exact}$ is the diagonal matrix of the exact energies, and $E$ is the diagonal matrix of the variational energies of the approximate eigenstates encapsulated in $U$.) As our approximate local integrals of motion $\tilde \tau_i^z$ do not commute with the Hamiltonian exactly, they do not persist to infinitely long times, but decohere on a timescale of the order of $1/\| [H, \tilde \tau_i^z\|_\mathrm{op}$~\cite{chandran2016higherD}. 
We note that, given the relatively small causal cones of our ansatz, the timescales we capture are shorter than the experimental ones. Therefore, absent other errors, the transition points we obtain will be lower bounds on the experimental ones (as for any disorder strength, localization might become unstable if we could consider longer timescales). Nonetheless, in our previous work using quantum circuits, we found excellent quantitative agreement between the experimental and theoretical transition point~\cite{2dMBLfermion}. 
We also note that, although atom loss is present in the experiment~\cite{cleanDirty2DMBL}, it is not relevant on the short time scales where our approximation is valid.

Once the optimized unitaries have been obtained, we can calculate one- and two-point functions of the approximate eigenstates (see Appendix~\ref{app:corfunc}). For the purpose of producing the phase diagram, we estimate the location of the transition (technically, crossover) point by using the fact that the variation of entanglement in the system between different disorder realizations is greatest at the critical disorder strength~\cite{kjall2014many}. Specifically, the process is as follows: one can extract the on-site entanglement entropies with respect to sites $i$. After averaging over eigenstates in particle sectors and calculating the fluctuation with respect to disorder realizations, one obtains maxima around the phase transition point~\cite{kjall2014many,Wahl2017PRX,Wahl2022,2dMBLfermion,Venn2023}. (To improve numerical stability, in the final step, one also averages the entanglement entropy fluctuation over sites $i$.) The disorder strengths $\Delta$ where these maxima are located can be used to extract the phase transition points. We carry out this analysis for overall filling fraction $\nu = 0.5$ as in the experiment~\cite{cleanDirty2DMBL} and as a function of the fraction $f_c \in [0, 1]$ of clean bosons, by explicitly selecting approximate eigenstates with the desired values of $\nu$ and $f_c$. For our simulations, as we consider the Hamiltonian as defined on a $10 \times 10$ lattice, we have $N_c + N_d = 50$ bosons.

\section{Results}
Here we present our findings on the phase diagram as a function of the fraction $f_c$ of clean particles along with a comparison to analogous results from simulation of real time evolution. We then show results on correlation functions and quantum mutual information. 

\subsection{Clean particle fraction-resolved phase diagram}

\begin{figure}[t]
\centering
\includegraphics[width=0.5 \textwidth]{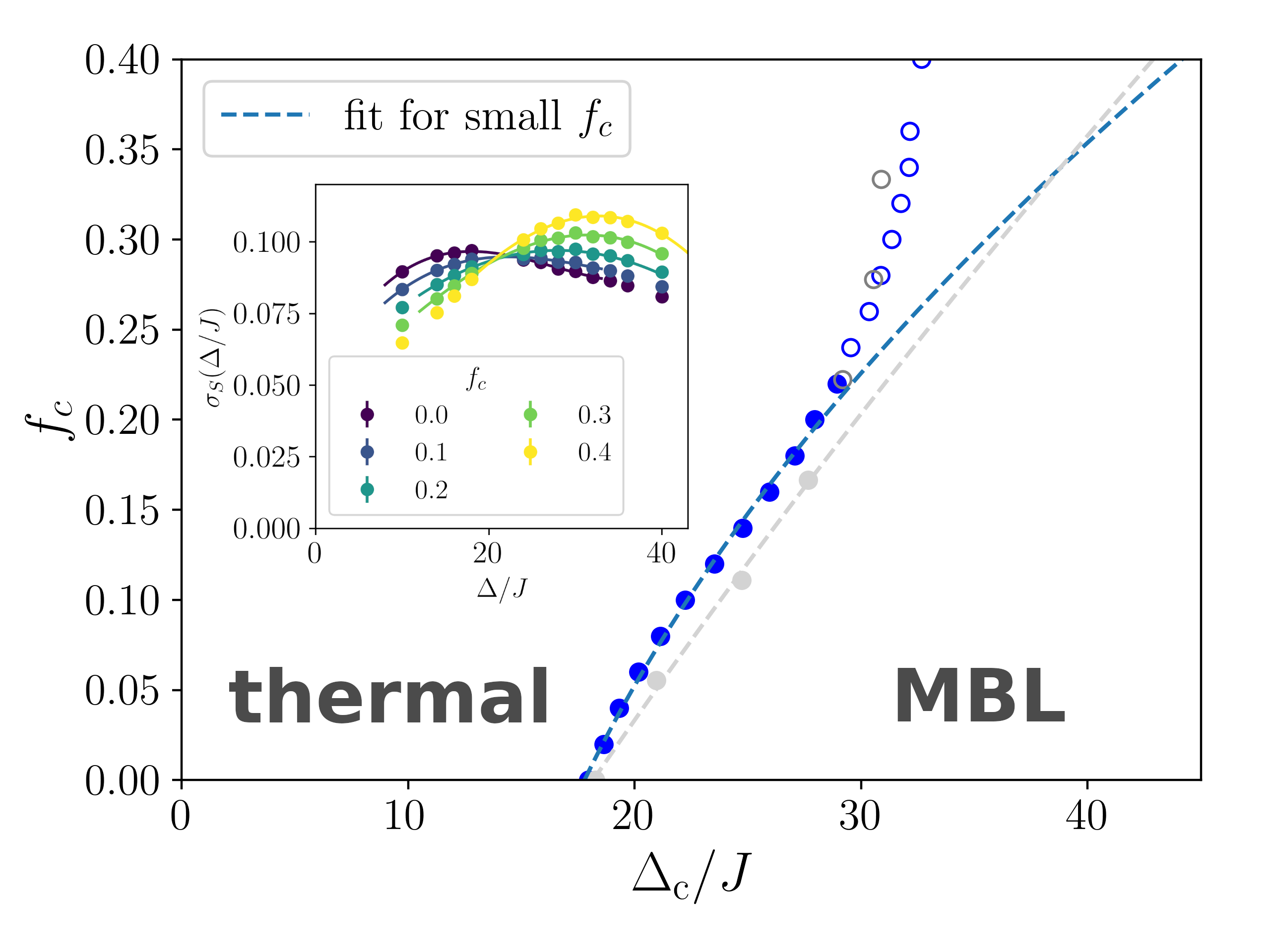}
    \caption{Phase diagram of thermal and MBL-like regimes for overall filling fraction $\nu = 0.5$ obtained by extracting the transition points from the maxima of the entanglement entropy fluctuation for different $f_c$ in the inset. The blue dashed line indicates a second order polynomial fit to our data for small $f_c$ (blue filled circles), from which our method deviates for $\Delta \gtrsim 30J$ (blue empty circles). The gray symbols indicate the corresponding results for a $6 \times 6$ system, see Appendix for details. 
    Inset: Entanglement entropy fluctuation (see main text) as a function of disorder strength $\Delta$ with fractions $f_c$ of clean particles as indicated in the legend. The error bars are smaller than the symbol size.} 
    \label{fig:phase_diagram}
\end{figure}

In Fig.~\ref{fig:phase_diagram} we show the numerical data obtained by optimizing our quantum circuit ansatz for 30 disorder realizations as a function of disorder strength $\Delta$. To increase numerical stability, we used the same 30 disorder profiles for all $\Delta$ and only varied the overall prefactor~\cite{Wahl2017PRX,2DMBL,2dMBLfermion}. As detailed above, we calculated the fluctuation of the eigenstate-averaged  (sampling over 4000 approximate eigenstates)
 on-site entanglement entropy with respect to disorder realizations and averaged the result over the site positions. We restricted the eigenstates to have $N_c + N_d \equiv N_\mr{tot} = 50$ bosons with $N_c/N_\mr{tot} = f_c$ fixed. 
 The corresponding filling fraction-resolved entanglement entropy fluctuation is shown in Fig.~\ref{fig:phase_diagram}. (We note that we did perform an error analysis, using the method described in Ref.~\cite{2dMBLfermion}---the resulting error bars are smaller than the scatter points and therefore not visible.) For all fractions $f_c$ of clean particles, we find a unique maximum in the entanglement entropy fluctuation (see inset of Fig.~\ref{fig:phase_diagram}), whose disorder strength value we use to plot the filling-fraction dependent MBL-thermal phase diagram. We note that for disorder strengths $\Delta \gtrsim 30J$ our phase boundary deviates from the expected behavior of approaching a horizontal line asymptotically.  Such a behavior is required, as the eigenstates for any arbitrarily large $\Delta$ are always delocalized if $f_c$ is close enough to 1, such that the $N_d = (1-f_c) N_\mr{tot}$ dirty bosons cannot localize the clean ones.  
 Finding the transition point for large disorder strengths gets increasingly difficult for our method, as there are fewer and fewer dirty particles, which have to localize the clean particles on larger and larger length scales in order to remain MBL. Our short-depth quantum circuit ansatz is unable to capture such large localization lengths and a breakdown of our approach for fractions of clean particles close to 1 is therefore expected. 
 
 Due to the asymptotic flattening of the phase boundary, the Taylor expansion of $f_c(\Delta_c)$ should have a negative coefficient in second order. We fit our data to such a second order polynomial in the regime where the data points follow a curve which is right-bent. Crucially, the transition point $\Delta_c(f_c=0) = 18J$ is consistent with Ref.~\cite{2DMBL}, i.e., the small $f_c$ results are reliable. Specifically, they match the expected (fitted) behavior in the range $0 \leq \Delta_c(f_c) \leq 30J$, which covers the value of $\Delta = 28J$ used in the experiment~\cite{cleanDirty2DMBL}. For the experimental disorder strength, we find a transition at $f_c (\Delta=28J) \approx 0.20$, which compares favorably with the experimental value of $f_c^\mr{exp} = N_c/N_\mr{tot} \approx (40\pm5)/(124\pm 12) \approx 0.33\pm 0.07$~\cite{cleanDirty2DMBL} given the simplicity of our approximation.  The still significant deviation from the experimental value is mostly due to the limited entanglement in the ansatz, and the truncation of the local Hilbert space. Specifically, the unitaries building up our quantum circuit act only on plaquettes of $2 \times 2$ sites, which limits the amount of entanglement allowed by our ansatz, and we considered a hardcore Bose-Hubbard model, i.e., neglected double and higher occupancies. Mitigating either of these restrictions would likely lead to a transition point closer to the experimental one, but would require vastly larger computational resources. (Our theoretical result of $f_c (\Delta=28J) \approx 0.20$ has negligible error bar, but the error analysis only takes into account statistical error due to disorder realizations and eigenstate sampling, not the systematic errors just mentioned.)

 We also note that the transition point $f_c$, from numerical simulations of a $10\times 10$ system, is closer to the experimental value than the one found for a $6 \times 6$ system, indicating the correct trend (cf. Appendix~\ref{app:6x6}). This underpins the statement that, due to their local structure, shallow quantum circuits are capable of capturing the transition points found with charge-density-wave imbalance experiments~\cite{2dMBLfermion}.

\subsection{Real time dynamics}\label{sec:tdvp}

Complementing our variational diagonalization, we also simulated time evolution using the time-dependent variational principle (TDVP) on matrix product states (MPS)~\cite{tdvp}, following the methods of Ref.~\cite{Elmer}. TDVP simulations were done using the TeNPy library~\cite{tenpy}. Specifically, we consider a $6 \times 6$ system with open boundary conditions and initialize it in a charge-density-wave state with alternating full and empty columns. The full columns are occupied with one boson per site -- randomly either a dirty or a clean atom -- such that the overall number of clean particles is $N_c$ and corresponding fraction of clean particles $f_c = N_c/18$. We performed the evolution under the hardcore Hamiltonian Eq.~\eqref{eq:Ham} and calculated the imbalance defined as $I(t) = \frac{N_e(t) - N_o(t)}{N_e(t) + N_o(t)}$ at each time step, where $N_{e/o}(t)$ is the total number of atoms in all even / odd columns. 
We found that for a maximum MPS bond dimension of $\chi = 384$, and time step size $\Delta t = 0.05$, the results for the imbalance are sufficiently well converged, even for long evolution times of $t = 100 J^{-1}$ hoppings. 

\begin{figure}[t]
\centering
\hspace{-12pt}\includegraphics[width=0.5 \textwidth]{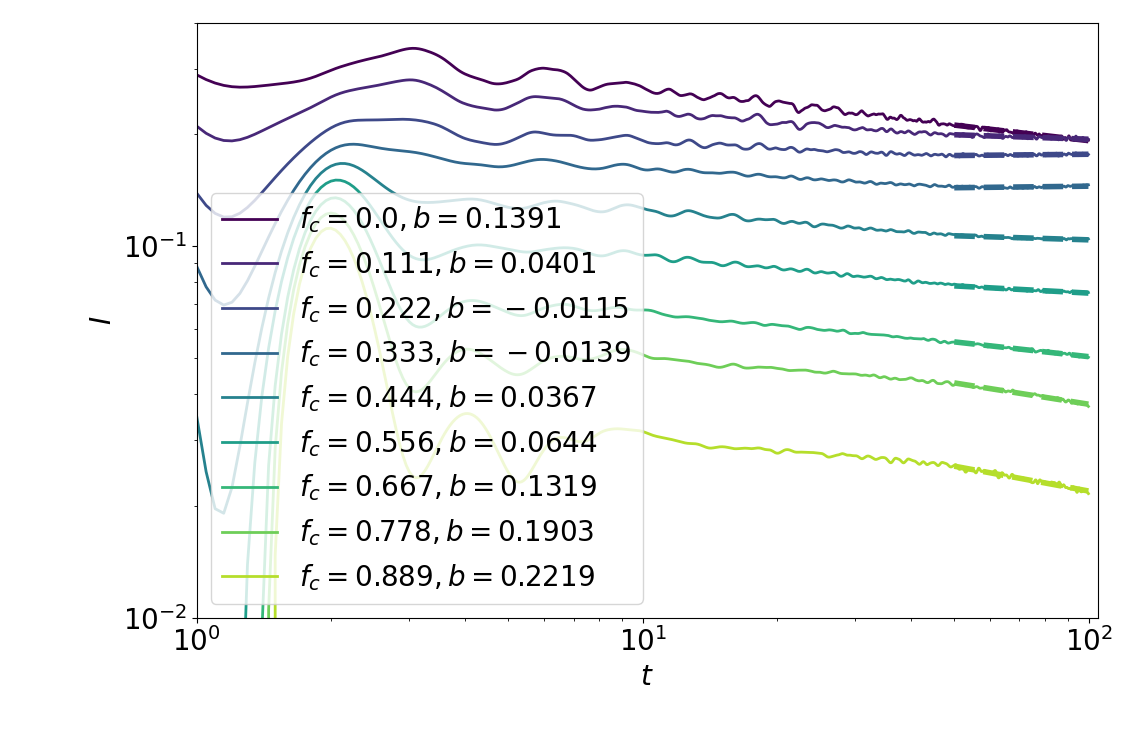}
    \caption{Imbalances $I$ as a function of time $t$ for different fractions $f_c = N_c/18$ of clean particles obtained for a $6 \times 6$ system initialized with a charge density wave of $N_c + N_d = 18$ bosons. The data are averaged over ten disorder realizations. The dashed lines indicate a polynomial fit $I = a t^{-b}$ for late times ($a,b \in \mathbb{R}$). Starting from large $f_c$, we find that the exponent $b$ vanishes at around $f_c \approx 0.4$. For small $f_c$, $b$ is positive again, as the initial exponential decay of the imbalance has not fully subsided yet.} 
    \label{fig:imbalances}
\end{figure}

\begin{figure}[t]
\centering
\includegraphics[width=0.5 \textwidth]{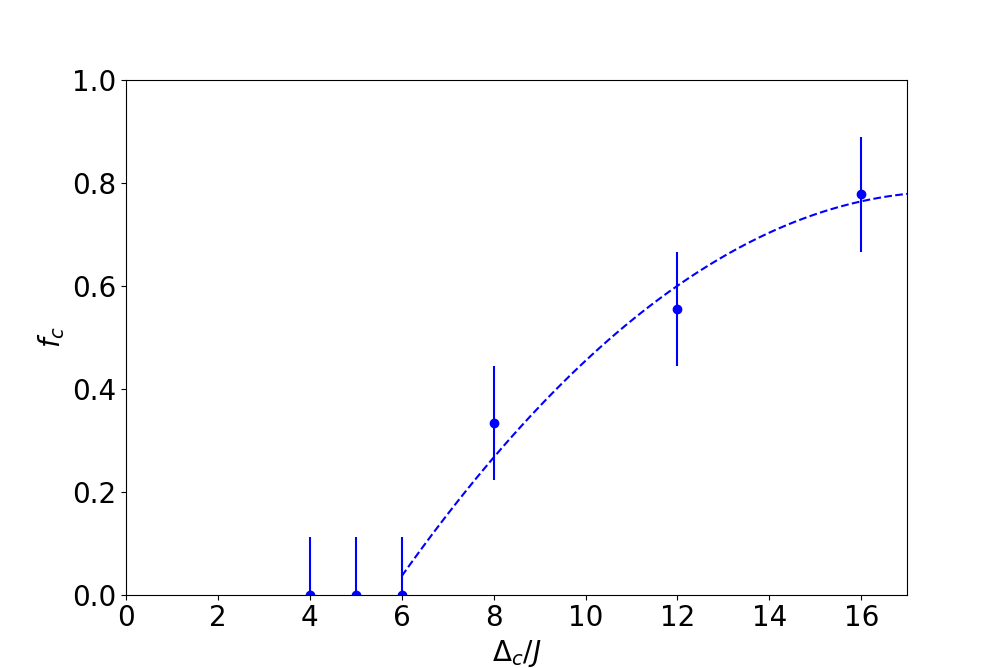}
    \caption{Phase diagram as a function of the fraction $f_c$ of clean particles and disorder strength $\Delta$ obtained from the imbalance evolutions for different values of $\Delta$ (cf. Fig.~\ref{fig:imbalances} for $\Delta = 8J$). We averaged over ten disorder realizations for each disorder strength. Error bars denote the uncertainty in determining at which fraction $f_c$ the exponent $b$ vanishes. The dashed line is a guide to the eye.} 
    \label{fig:phase_diagram_TDVP}
\end{figure}

We present the results for the imbalance as a function of time and for the example of disorder strength $\Delta = 8J$ and different  fractions $f_c$ of clean particles in Fig.~\ref{fig:imbalances}. The dashed lines in the plots indicate polynomial fits whose exponents can be used to determine the critical value of $f_c$ for a given disorder strength (as the fraction $f_c$ where the exponent of the polynomial fit becomes numerically zero)~\cite{Elmer}. Afterwards, we used the critical fractions $f_c$ for different disorder strengths to infer the filling fraction-dependent phase diagram, shown in Fig.~\ref{fig:phase_diagram_TDVP}. Compared with Fig.~\ref{fig:phase_diagram}, we notice that the critical disorder strengths for fixed $f_c$ are shifted to significantly lower values, incompatible with the experimental transition point of $f_c^\mr{exp} \approx 0.33 \pm 0.07$. We ascribe this to strong boundary effects (20 sites of 36 are next to the edge) and the two-dimensional MBL-to-thermal transition points being at significantly lower disorder strengths for smaller systems~\cite{Gopalakrishnan2019,Elmer}. However, we note that the phase diagram obtained with TDVP is qualitatively similar to the one obtained with quantum circuits. In the former case, we are unfortunately unable to reach much larger system sizes due to limitations of computational resources. (The bond dimensions required for sufficient accuracy quickly get too large.) Furthermore, closing the boundaries would lead to additional, spurious delocalization not present in the experimental system. We thus conclude that quantum circuit simulations are better suited than TDVP to reproduce the two-dimensional MBL-to-thermal transition points of the considered two-species Bose-Hubbard Hamiltonian, at least for experimentally relevant system sizes. Nonetheless, the TDVP simulations qualitatively support our quantum circuit results.

\subsection{Two-point correlation function}
We now return to the discussion of results obtained from the quantum circuit study.  Our method allows us to use the optimized unitaries to calculate two-site reduced density matrices (see Appendix~\ref{app:corfunc} for details), from which we can compute two-point correlation functions and the quantum mutual information (QMI) between sites.

\begin{figure}[t]
\centering
\includegraphics[width=0.5 \textwidth]{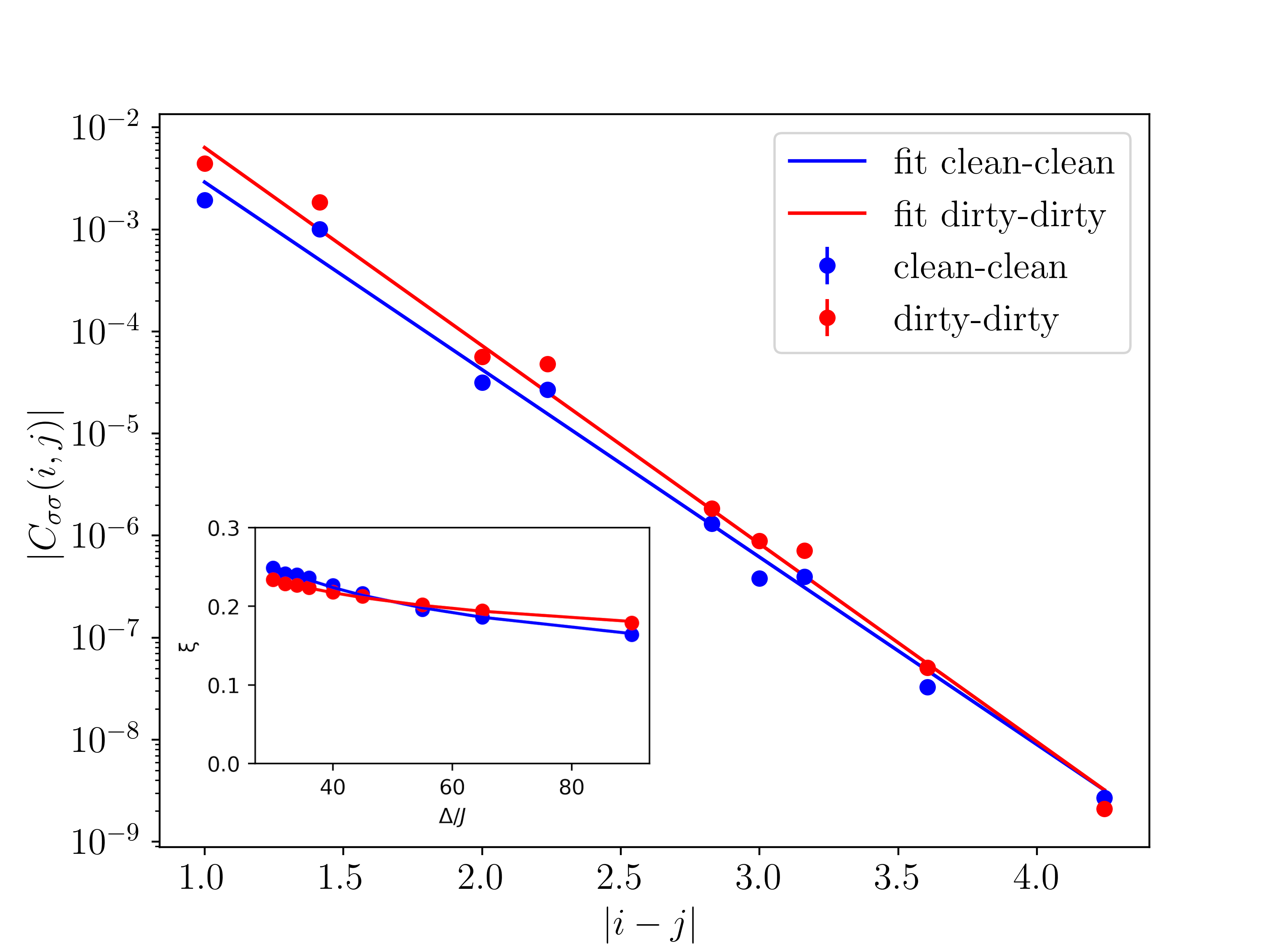}
    \caption{Decay of correlations as a function of distance $|i -j|$, shown for the example of $\Delta = 36J$. The data were obtained by averaging over 100 randomly chosen approximate eigenstates with fraction $f_c = 0.2$ of clean bosons and averaging over 30 disorder realizations. Error bars denote fluctuations between disorder realizations and are smaller than the symbol size.  
    For the fits, we used linear functions after taking the logarithm for the three data points of smallest distance. Inset: Correlation lengths extracted from the fits as shown in the main figure for various values of $\Delta$ and $f_c = 0.2$. The solid lines denote $1/\ln(\Delta/J)$ fits~\cite{Hundertmark,Pietracaprina2016,Scardicchio2017}.}  \label{fig:correlations}
\end{figure}

We first consider correlation functions. Within our approximation, the two-site correlation function $C_{\sigma \mu}(i,j) =  \ev{n_{i,\sigma} n_{j, \mu}} - \ev{n_{i,\sigma}} \ev{n_{j,\mu}}$, with $\sigma, \mu \in \{c,d\}$, is non-trivial only in a fixed range of separations between sites $i$ and $j$ (causal) cone. The obtained correlations and extracted correlation lengths are shown in Fig.~\ref{fig:correlations}. We note the slow decay of the average correlation lengths with disorder strength $\Delta$, cf. Ref.~\cite{2DMBL}. Furthermore, we find that the correlation lengths for dirty and clean particles are very similar, indicating that the dirty particles are effective at localizing the clean ones in the MBL regime within our approximation. 
For example, for $f_c = 0.2$ we observe that $\overline{\xi_{cc}}(\Delta=45J) = 0.22\pm0.01$ and $\overline{\xi_{dd}}(\Delta=45J) = 0.21\pm0.01$, decaying to $\overline{\xi_{cc}}(\Delta=90J) = 0.16\pm0.01$ and $\overline{\xi_{dd}}(\Delta=90J) = 0.18\pm0.01$. Our predictions for the two-point correlations can be compared to future optical lattice experiments. 

\subsection{Quantum mutual information}
Finally, we consider the quantum mutual information (QMI), which has been used to quantify correlations in MBL systems~\cite{deTomasiQMI, villalonga2020characterizing}, and more recently has been studied in the context of resonances~\cite{Morningstar2022}.  Here, we analyze the distribution of the QMI between individual sites in our two-dimensional model \eqref{eq:Ham}. The QMI of an eigenstate $|\psi\rangle$ is defined as
\begin{align}
I(A:B) = S(\rho_A) + S(\rho_B) - S(\rho_{AB}),
\end{align}
where the system is partitioned into three disjoint parts $A$, $B$, $C$, and the QMI is between regions $A$ and $B$. $\rho_X$ is the reduced density matrix of $|\psi\rangle$ obtained by tracing out its complement $\overline X$,
\begin{align}
\rho_X = \trace_{\overline{X}} (|\psi\rangle\langle\psi|),
\end{align}
and $S(\rho_X) = -\trace(\rho_X \ln (\rho_X))$ denotes the von Neumann entropy.

\begin{figure*}
    \centering
\includegraphics[width= \textwidth]{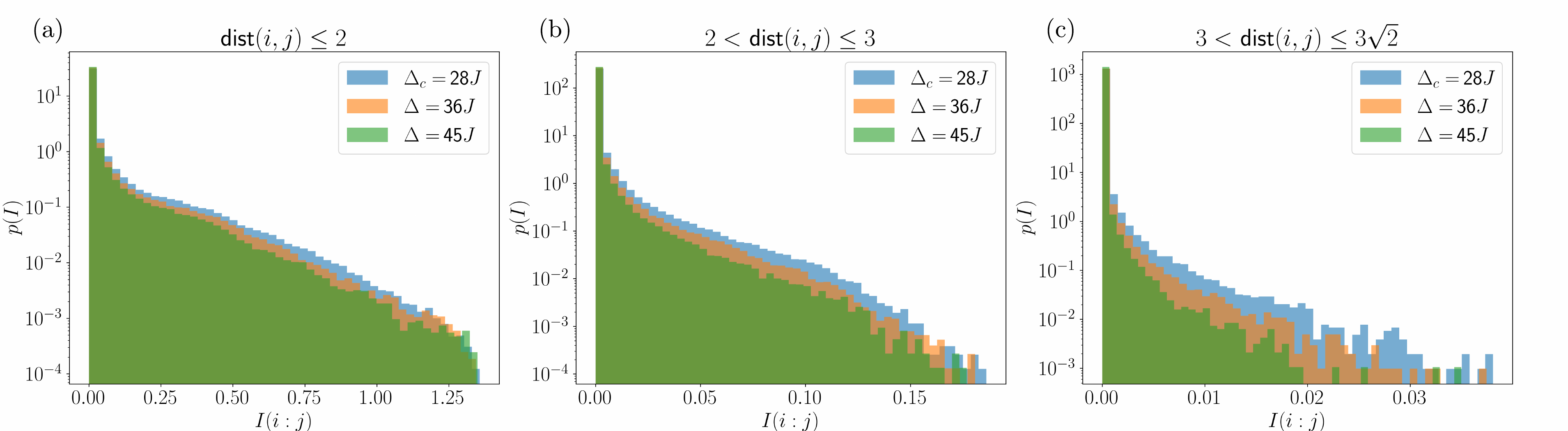} 
    \caption{Probability density of the two-site QMI obtained by sampling over 100 eigenstates and 30 disorder realizations, taking all sites into account, for $f_c = 0.2$ (corresponding to a transition point $\Delta_c = 28J$) and the values of $\Delta$ indicated in the legend. The distances between the sites are $\mr{dist}(i,j) \leq 2$ in (a), $2 < \mr{dist}(i,j) \leq 3$ in (b), and $3 < \mr{dist}(i,j)\leq 3\sqrt{2}$ in (c).} 
    \label{fig:QMI}
\end{figure*}

We first review the expected behavior of the eigenstate QMI probability distribution.  In general, the two-site QMI $I(i:j)$ decays exponentially with distance in the MBL regime, and decays more slowly in the thermal regime~\cite{deTomasiQMI}. The probability distribution of values of the two-site QMI in thermal-regime eigenstates is roughly symmetric and centers around a value independent of the distance $|i-j|$, while in the MBL regime, the QMI distribution is more one-sided with greatest probability near zero, but also longer tailed, with more instances of near-maximal QMI, particularly for small $|i-j|$~\cite{villalonga2020characterizing}.  When considering the QMI distributions at various disorder strengths, the distribution is expected to be independent of disorder strength for neighboring sites, while for large $|i-j|$ the distribution is more heavily centered around $0$ at very strong disorder, and longer tailed closer to the transition point. 

Using our variational method, we calculate the QMI, from the two-site reduced density matrices (which can be obtained as illustrated in Appendix~\ref{app:corfunc}), sampling over 100 approximate eigenstates per disorder realization and disorder strength. As our ansatz only captures area law-entangled states, we are unable to study the QMI distribution in the thermal regime. The results, for disorder strengths at the transition point and above, are shown in Fig.~\ref{fig:QMI}. We note that although the local Hilbert space dimension of this model is $3$, the maximum QMI between two sites is $I(i:j) = 2\ln2$, equivalent to that of two maximally entangled qubits, as particle number conservation prohibits the existence of maximally entangled two-qutrit states on a subsystem of two sites.  

As our approximation only captures non-trivial correlations within a fixed range of distances, we can only consider certain separations between sites $i$ and $j$. We consider the short-ranged, i.e. nearest-neighbor case in Fig.~\ref{fig:QMI}(a), where the distributions are roughly similar for different disorder strengths and show instances of near-maximal QMI, as expected based on analogous studies in one-dimensional MBL. We also consider the ``medium-ranged" case in Fig.~\ref{fig:QMI}(b) and (c).  We are unable to consider the long-ranged case, as, within our approximation, sites are uncorrelated for $|i-j|>3\sqrt{2}$. In addition, a further limitation arising from the local nature of the ansatz is that the QMI is consistently $\lesssim 0.2$ for $|i-j|\geq2$ and $\lesssim 0.05$ for $|i-j|>3$, independently of the model under consideration. Thus in Fig.~\ref{fig:QMI}(b) and (c), the observed values of the QMI are likely significantly below the true values.  However, the qualitative aspects of these plots may still capture some of the relevant physics.  This is because each approximate eigenstate is essentially the projection of an exact eigenstate onto a manifold of weakly entangled states; since the projection is the same in all cases, it is still meaningful to consider \textit{relative} differences, especially correlations or related quantities between different sets of eigenstates.  In particular, this allows us to compare the two-site QMI, at fixed distances, between systems at different disorder strengths.

We now discuss the relation to resonances, which have been suggested to be the driving force behind the transition from the MBL to the thermal regime in one dimension~\cite{Khemani2017MBLT,Villalonga2020,Morningstar2022,Crowley2022,Ha2023}. Resonances can be viewed as superpositions of few (deep in the MBL regime, two)  near-product states whose local occupations differ in many sites, e.g., cat states. A crucial requirement is that those superpositions are the result of the near-product states having hybridized due to them being close in energy. This can be detected via the QMI if the regions $A$ and $B$ (in our case, sites $i$ and $j$) are far away (typically of the order of the system size), such that a non-vanishing QMI indicates the superposition of macroscopically distinct near-product states, which, due to the locality of the Hamiltonian, must be close in energy~\cite{Morningstar2022}. Due to the locality of our ansatz, we are unable to detect such long-range resonances. Instead, we study how the QMI between the most distant sites accessible builds up as the disorder strength $\Delta$ is lowered towards the transition. 

We observe that in Fig.~\ref{fig:QMI}(c) the distribution of the QMI develops tails near the transition that are absent for larger $\Delta$. We note that the lengths of these tails are constrained due to the aforementioned limitations arising from the short-range nature of the ansatz.  We do not observe the near-maximal QMI that is the hallmark of many-body resonances, in the ``medium-ranged" case.  However, the qualitative behavior of the distribution is reminiscent of the expected behavior of the distribution of the longer-range QMI as the disorder strength is lowered towards the transition point.  Although we do not capture resonances with our numerical methods, within the range of QMI and site separations that we do capture, we observe a broadening in the QMI distribution with decreasing $\Delta$ that corresponds to the increased presence of resonances near the transition point, analogous to the one-dimensional case.

\section{Conclusions}

\begin{figure*}
    \centering
    \includegraphics[width=0.6\linewidth]{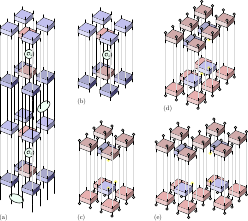}
    \caption{(a) Illustration of the summands that appear in $\Tr((H \tilde \tau_{i,\mu})^2)$ in the cost function given in Eq.~\eqref{eq:cost_fn}, where the ovals illustrate nearest-neighbour terms in $H$. Note that the trace would correspond to the open legs at the top and the bottom being identified.
    (b) Illustration of $U^\dagger\mathcal{O}_iU$ of a local operator  $\mathcal{O}_i$ acting on site $i$. Using unitarity, $U^\dagger\mathcal{O}_iU$ has non-trivial support only in the causal cone of operator $\mathcal{O}_i$.
    (c-e) Diagrammatic representations of the two-site reduced density matrix $\rho_{ij} = \Tr_{\overline{ij}}(\ketbra{\psi})$. Different choices of the sites $i$ and $j$ lead to causal cones of size (c) $4\times 4$, (d) $4\times 6$, and (e) 28 sites. Note that these diagrams are inverted: the top tensor network is $\bra{\psi}$ and the bottom tensor network is $\ket{\psi}$. For clarity, the tensor contractions are shown with light gray lines, and the gray circles pinned to the outer legs indicate that these indices are fixed to particular values and hence no longer open. The open tensor legs (on sites $i$ and $j$) are highlighted in yellow. Note that here we depict the case where the distance $|i-j|$ is $1$, $1$, and $\sqrt{2}$ in (c), (d), and (e) respectively, but different choices of $i$ and $j$ lead to the same types of tensor contractions.  In particular, $|i-j|$ can be up to $\sqrt{2}$ in case (c), $\sqrt{10}$ in case (d), and $3\sqrt{2}$ in case (e).}
    \label{fig:rhoCalculations}
\end{figure*}

We used quantum circuits to simulate a two-component ultracold bosonic gas, where one component is localized by a disorder potential and the other acts as a heat bath. We found a transition point of the fraction of clean particles in the same regime as the experiments of Ref.~\cite{cleanDirty2DMBL}. We also calculated the correlation lengths as quantities to be compared to future experiments. Finally, we computed the QMI of our approximate eigenstates; 
we found that the distribution of the QMI gets broader as the disorder strength is decreased towards the transition. 
This corresponds to the perspective that the proliferation of resonances contributes to the two-dimensional MBL-to-thermal crossover seen on experimental time scales -- analogous to the case in one dimension.
Further studies are required to confirm this hypothesis.  

\begin{acknowledgments} 
We thank David Huse and Hannes Pichler for helpful discussions. JL acknowledges support from ERC Starting grant QARA (Grant No.~101041435) and the Austrian Science Fund (FWF): COE 1 and quantA. AC acknowledges support from the Royal Society grant RGS{$\backslash$}R1{$\backslash$}231444 and from the EPSRC Open Fellowship EP/X042812/1. TBW acknowledges funding from an EPSRC ERC underwrite grant EP/X025829/1. The computational results presented here have been achieved (in part) using the LEO HPC infrastructure of the University of Innsbruck.
\end{acknowledgments}

\appendix

\section{Variational ansatz}\label{app:sec1}
In this work, we use the quantum circuit $U$, a two-layer tensor network with the structure depicted in Fig.~\ref{fig:combined}(c), to variationally diagonalize the Hamiltonian \eqref{eq:Ham}, i.e. approximately calculate the full set of eigenstates.  (More precisely, only a subset of the tensor network is depicted in Fig.~\ref{fig:combined}(c), for clarity.  Since we consider a $10\times10$ system in our numerics, both the lower layer and upper layer consist of $5^2$ tensors, with the tensors on the edges connected as prescribed by periodic boundary conditions.)

$U$ may also be recast as a large matrix in the full Hilbert space; in this case, the columns of $U$ are the approximate eigenstates.  In the tensor network picture, we adopt the convention where the upper indices are the row indices (or contravariant/``ket" indices) and the lower indices are the column indices (or covariant/``bra" indices).  Thus, to extract a particular approximate eigenstate from $U$, one fixes all the lower indices to particular values: the resulting tensor network has only upper open indices and is a projected entangled pair state (PEPS). The indices take the values $\{0,c,d\}$, corresponding to the states $\ket{0},\ket{c}$, and $\ket{d}$, which span the single-site Hilbert space.  

Crucially, $U$ is unitary and also conserves particle number and type, i.e. each approximate eigenstate is of a definite number of clean and dirty particles. In our numerics, these conditions are enforced on the individual tensors in the network [boxes in Fig.~\ref{fig:combined}(c)], which we call $u$. (Note that, in general, all tensors are different, as the system is not translationally invariant; here $u$ refers to a generic one of them.)  

To impose particle number conservation, each $u$ is subject to the condition that the element $ u^{\alpha,\beta,\gamma,\delta}_{\alpha',\beta',\gamma',\delta'}$ is nonzero only if ${\alpha,\beta,\gamma,\delta}={\alpha',\beta',\gamma',\delta'}$ up to permutation.  That is, the four upper indices of $u$ must contain the same number of both $c$'s and $d$'s as the four lower indices. 
 $u$ is also unitary, or more specifically, real and orthogonal---it suffices to have a real ansatz since the Hamiltonian is real. This is imposed by setting $u = e^M$, where $M$ is a real anti-symmetric matrix that contains the actual variational parameters.

The quantum circuit structure of $U$ ensures that calculations involving local operators are local tensor network contractions of finite range, independent of the overall system size.  For example, consider the quantity $U^\dagger\mathcal{O}_iU$, where $\mathcal{O}_i$ is a single-site operator acting on site $i$.  On sites away from $i$, the $u$ and $u^\dagger$ tensors will cancel due to unitarity. Thus, to calculate $U^\dagger\mathcal{O}_iU$, one only needs to consider a $4\times 4$ causal cone of the local operator $\mathcal{O}_i$ as depicted in Fig.~\ref{fig:rhoCalculations}(b).

\section{Reduced density matrix calculations}\label{app:corfunc}
Here we describe how we calculate the one- and two-site reduced density matrices from the approximate eigenstates in the optimized $U$ and how we then use the density matrices to obtain the correlation functions and correlation lengths. 

Let $\ket{\psi}$ be a (randomly selected) approximate eigenstate. The setup of our ansatz as a unitary quantum circuit allows us to obtain the single-site density matrix $\rho_i = \Tr_{\overline{i}}(\op{\psi})$ and two-site density matrix $\rho_{ij} = \Tr_{\overline{ij}}(\ketbra{\psi})$ via causal cone calculations. However, the two-site density matrix trivially decomposes as $\rho_{ij} = \rho_i\otimes\rho_j$ unless sites $i$ and $j$ have overlapping causal cones, which implies that one site must lie in a $6\times 6$ region around the other.  Thus, for each site $i$ we consider all $35$ other sites $j$ for which nontrivial correlations can be calculated.  In Fig.~\ref{fig:rhoCalculations} we illustrate the tensor network contractions required to calculate $\rho_{ij}$ in various cases. The calculation of $\rho_i$ involves a diagram that is the same as Fig.~\ref{fig:rhoCalculations}(d), but with open legs on only one site $i$.

Due to the square lattice geometry, the distances considered are (in units of the lattice spacing) 
\begin{align}
    |i-j| &= 1, \sqrt{2}, 2, \sqrt{5}, 2\sqrt{2}, 3, \sqrt{10}, \sqrt{13}, 3\sqrt{2} \notag \\
    &\approx 1, 1.41, 2, 2.23, 2.83, 3, 3.16, 3.61, 4.24.
\end{align}

From the single-site reduced density matrices, we can calculate the on-site entanglement entropies $S_i = -\Tr(\rho_i\ln\rho_i)$. From-the two site density matrices, we can calculate $S_{ij} = -\Tr(\rho_{ij}\ln\rho_{ij})$ and thus the quantum mutual information (QMI) between sites. We also calculate the two-point correlation functions, which are defined as   
\begin{equation}
   C_{\sigma \mu}(i,j) =  \ev{n_{i,\sigma} n_{j, \mu}} - \ev{n_{i,\sigma}} \ev{n_{j,\mu}},
\end{equation}
where $\langle\cdots\rangle=\ev{\cdots}{\psi}$, so $\langle n_{i,\sigma}\rangle =  \Tr(\rho_i n_{\sigma})$ and $\langle n_{i,\sigma}n_{j,\mu}\rangle  = \Tr(\rho_{ij} (n_{\sigma}\otimes n_{\mu}))$, and where $\sigma,\mu\in\{c,d\}$.  We then calculate $C_{\sigma\mu}(i,j)$ at various points and perform an exponential fit $C_{\sigma\mu}(i,j)\propto\exp(-|i-j|/\xi_{\sigma\mu})$ to obtain the correlation length $\xi_{\sigma\mu}$.  Note that here $C_{\sigma\mu}$ and hence $\xi_{\sigma\mu}$ corresponds to a specific eigenstate $|\psi\rangle$. We average over eigenstates and disorder realizations to obtain the average quantities $\overline{C_{\sigma\mu}}(\Delta)$ and $\overline{\xi_{\sigma\mu}}(\Delta)$ at a given disorder strength $\Delta$.

\begin{figure}[t]
\centering
\includegraphics[width=0.5\textwidth]{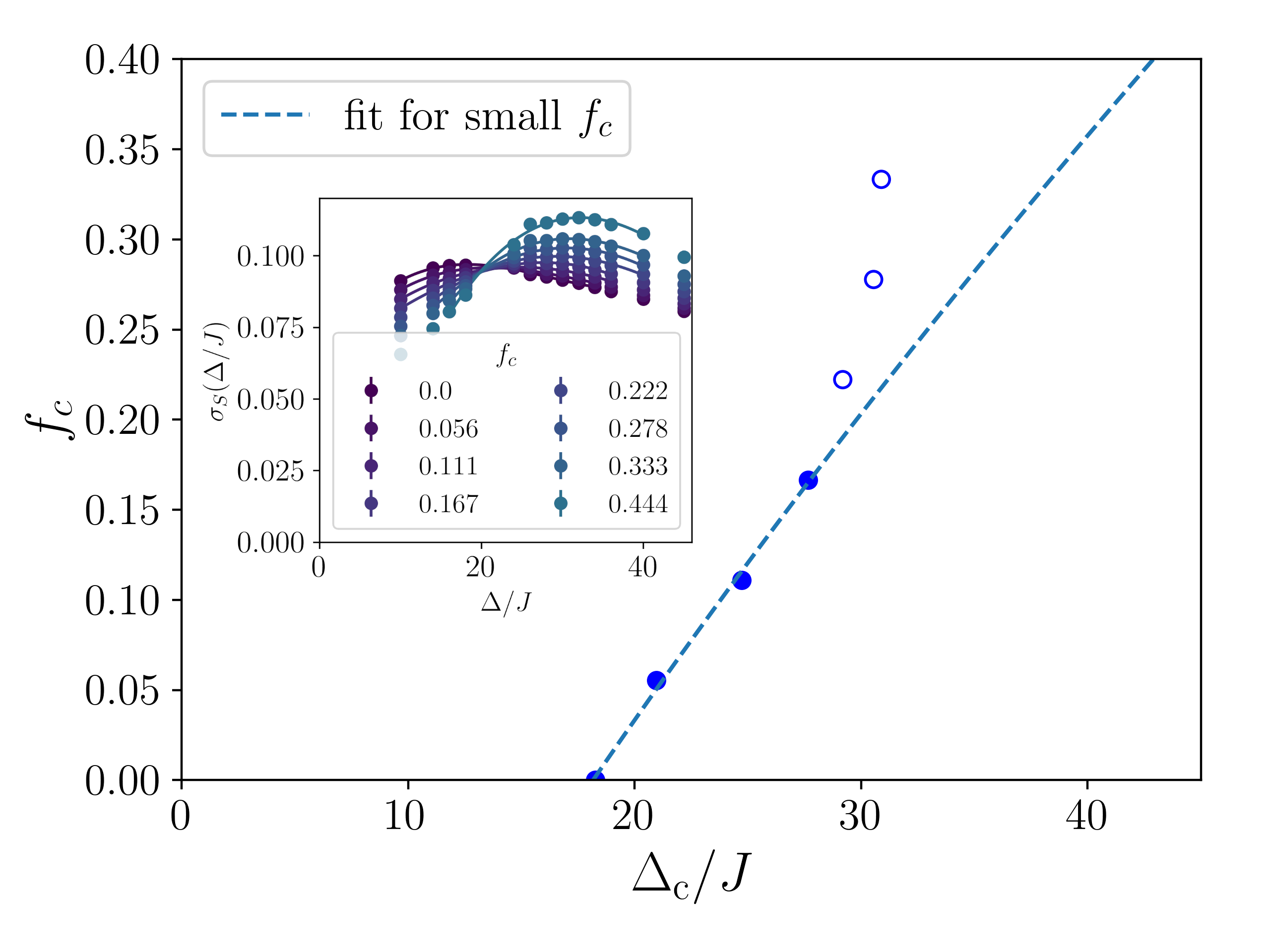}

    \caption{Clean-fraction resolved phase diagram and (inset) entropy fluctuations, for a $6\times 6$ system, for overall filling fraction $\nu = 0.5$, analogous to Fig.~\ref{fig:phase_diagram} in the main text.} 
    \label{fig:6x6}
\end{figure}

\section{$6\times 6$ results}\label{app:6x6}
To verify the convergence of the results with increasing system size, we performed a set of calculations for a $6\times 6$ system, analogous to the $10\times 10$ calculations  described in the main text. The results, i.e., the $6\times 6$ analog of Fig.~\ref{fig:phase_diagram}, are shown in Fig.~\ref{fig:6x6}.  For the disorder strength used in the experiment, we find that the transition occurs at $f_c(\Delta=28J)=0.17$, which is somewhat lower than the value $0.20$ obtained from the $10\times 10$ simulations. 

The results are  similar to those of the $10\times 10$ system, although with fewer data points. We note that our method does not have any significant finite-size effects, but instead has finite-entanglement effects, as given by the finite range of the causal cones and correspondingly the finite range of correlations our approximation can give rise to. The causal cones of our approximate local integrals of motion have a finite range, and as a result they essentially describe the bulk of a two-dimensional infinitely large MBL system.

\bibliography{references}

\begin{thebibliography}{53}%
\makeatletter
\providecommand \@ifxundefined [1]{%
 \@ifx{#1\undefined}
}%
\providecommand \@ifnum [1]{%
 \ifnum #1\expandafter \@firstoftwo
 \else \expandafter \@secondoftwo
 \fi
}%
\providecommand \@ifx [1]{%
 \ifx #1\expandafter \@firstoftwo
 \else \expandafter \@secondoftwo
 \fi
}%
\providecommand \natexlab [1]{#1}%
\providecommand \enquote  [1]{``#1''}%
\providecommand \bibnamefont  [1]{#1}%
\providecommand \bibfnamefont [1]{#1}%
\providecommand \citenamefont [1]{#1}%
\providecommand \href@noop [0]{\@secondoftwo}%
\providecommand \href [0]{\begingroup \@sanitize@url \@href}%
\providecommand \@href[1]{\@@startlink{#1}\@@href}%
\providecommand \@@href[1]{\endgroup#1\@@endlink}%
\providecommand \@sanitize@url [0]{\catcode `\\12\catcode `\$12\catcode `\&12\catcode `\#12\catcode `\^12\catcode `\_12\catcode `\%12\relax}%
\providecommand \@@startlink[1]{}%
\providecommand \@@endlink[0]{}%
\providecommand \url  [0]{\begingroup\@sanitize@url \@url }%
\providecommand \@url [1]{\endgroup\@href {#1}{\urlprefix }}%
\providecommand \urlprefix  [0]{URL }%
\providecommand \Eprint [0]{\href }%
\providecommand \doibase [0]{https://doi.org/}%
\providecommand \selectlanguage [0]{\@gobble}%
\providecommand \bibinfo  [0]{\@secondoftwo}%
\providecommand \bibfield  [0]{\@secondoftwo}%
\providecommand \translation [1]{[#1]}%
\providecommand \BibitemOpen [0]{}%
\providecommand \bibitemStop [0]{}%
\providecommand \bibitemNoStop [0]{.\EOS\space}%
\providecommand \EOS [0]{\spacefactor3000\relax}%
\providecommand \BibitemShut  [1]{\csname bibitem#1\endcsname}%
\let\auto@bib@innerbib\@empty
\bibitem [{\citenamefont {Gornyi}\ \emph {et~al.}(2005)\citenamefont {Gornyi}, \citenamefont {Mirlin},\ and\ \citenamefont {Polyakov}}]{gornyi2005interacting}%
  \BibitemOpen
  \bibfield  {author} {\bibinfo {author} {\bibfnamefont {I.~V.}\ \bibnamefont {Gornyi}}, \bibinfo {author} {\bibfnamefont {A.~D.}\ \bibnamefont {Mirlin}},\ and\ \bibinfo {author} {\bibfnamefont {D.~G.}\ \bibnamefont {Polyakov}},\ }\bibfield  {title} {\bibinfo {title} {{Interacting electrons in disordered wires: Anderson localization and low-T transport}},\ }\href {https://doi.org/10.1103/PhysRevLett.95.206603} {\bibfield  {journal} {\bibinfo  {journal} {Phys. Rev. Lett.}\ }\textbf {\bibinfo {volume} {95}},\ \bibinfo {pages} {206603} (\bibinfo {year} {2005})}\BibitemShut {NoStop}%
\bibitem [{\citenamefont {Basko}\ \emph {et~al.}(2006)\citenamefont {Basko}, \citenamefont {Aleiner},\ and\ \citenamefont {Altshuler}}]{basko2006metal}%
  \BibitemOpen
  \bibfield  {author} {\bibinfo {author} {\bibfnamefont {D.~M.}\ \bibnamefont {Basko}}, \bibinfo {author} {\bibfnamefont {I.~L.}\ \bibnamefont {Aleiner}},\ and\ \bibinfo {author} {\bibfnamefont {B.~L.}\ \bibnamefont {Altshuler}},\ }\bibfield  {title} {\bibinfo {title} {{Metal--insulator transition in a weakly interacting many-electron system with localized single-particle states}},\ }\href {https://doi.org/10.1016/j.aop.2005.11.014} {\bibfield  {journal} {\bibinfo  {journal} {Ann. Phys.}\ }\textbf {\bibinfo {volume} {321}},\ \bibinfo {pages} {1126} (\bibinfo {year} {2006})}\BibitemShut {NoStop}%
\bibitem [{\citenamefont {\v{Z}nidari\v{c}}\ \emph {et~al.}(2008)\citenamefont {\v{Z}nidari\v{c}}, \citenamefont {Prosen},\ and\ \citenamefont {Prelov\v{s}ek}}]{znidaric2008many}%
  \BibitemOpen
  \bibfield  {author} {\bibinfo {author} {\bibfnamefont {M.}~\bibnamefont {\v{Z}nidari\v{c}}}, \bibinfo {author} {\bibfnamefont {T.}~\bibnamefont {Prosen}},\ and\ \bibinfo {author} {\bibfnamefont {P.}~\bibnamefont {Prelov\v{s}ek}},\ }\bibfield  {title} {\bibinfo {title} {{Many-body localization in the Heisenberg XXZ magnet in a random field}},\ }\href {https://doi.org/10.1103/PhysRevB.77.064426} {\bibfield  {journal} {\bibinfo  {journal} {Phys. Rev. B}\ }\textbf {\bibinfo {volume} {77}},\ \bibinfo {pages} {064426} (\bibinfo {year} {2008})}\BibitemShut {NoStop}%
\bibitem [{\citenamefont {Pal}\ and\ \citenamefont {Huse}(2010)}]{pal2010mb}%
  \BibitemOpen
  \bibfield  {author} {\bibinfo {author} {\bibfnamefont {A.}~\bibnamefont {Pal}}\ and\ \bibinfo {author} {\bibfnamefont {D.~A.}\ \bibnamefont {Huse}},\ }\bibfield  {title} {\bibinfo {title} {Many-body localization phase transition},\ }\href {https://doi.org/10.1103/PhysRevB.82.174411} {\bibfield  {journal} {\bibinfo  {journal} {Phys. Rev. B}\ }\textbf {\bibinfo {volume} {82}},\ \bibinfo {pages} {174411} (\bibinfo {year} {2010})}\BibitemShut {NoStop}%
\bibitem [{\citenamefont {Schreiber}\ \emph {et~al.}(2015)\citenamefont {Schreiber}, \citenamefont {Hodgman}, \citenamefont {Bordia}, \citenamefont {L{\"u}schen}, \citenamefont {Fischer}, \citenamefont {Vosk}, \citenamefont {Altman}, \citenamefont {Schneider},\ and\ \citenamefont {Bloch}}]{Schreiber842}%
  \BibitemOpen
  \bibfield  {author} {\bibinfo {author} {\bibfnamefont {M.}~\bibnamefont {Schreiber}}, \bibinfo {author} {\bibfnamefont {S.~S.}\ \bibnamefont {Hodgman}}, \bibinfo {author} {\bibfnamefont {P.}~\bibnamefont {Bordia}}, \bibinfo {author} {\bibfnamefont {H.~P.}\ \bibnamefont {L{\"u}schen}}, \bibinfo {author} {\bibfnamefont {M.~H.}\ \bibnamefont {Fischer}}, \bibinfo {author} {\bibfnamefont {R.}~\bibnamefont {Vosk}}, \bibinfo {author} {\bibfnamefont {E.}~\bibnamefont {Altman}}, \bibinfo {author} {\bibfnamefont {U.}~\bibnamefont {Schneider}},\ and\ \bibinfo {author} {\bibfnamefont {I.}~\bibnamefont {Bloch}},\ }\bibfield  {title} {\bibinfo {title} {Observation of many-body localization of interacting fermions in a quasirandom optical lattice},\ }\href {https://doi.org/10.1126/science.aaa7432} {\bibfield  {journal} {\bibinfo  {journal} {Science}\ }\textbf {\bibinfo {volume} {349}},\ \bibinfo {pages} {842} (\bibinfo {year} {2015})}\BibitemShut {NoStop}%
\bibitem [{\citenamefont {Smith}\ \emph {et~al.}(2016)\citenamefont {Smith}, \citenamefont {Lee}, \citenamefont {Richerme}, \citenamefont {Neyenhuis}, \citenamefont {Hess}, \citenamefont {Hauke}, \citenamefont {Heyl}, \citenamefont {Huse},\ and\ \citenamefont {Monroe}}]{Smith_MBL}%
  \BibitemOpen
  \bibfield  {author} {\bibinfo {author} {\bibfnamefont {J.}~\bibnamefont {Smith}}, \bibinfo {author} {\bibfnamefont {A.}~\bibnamefont {Lee}}, \bibinfo {author} {\bibfnamefont {P.}~\bibnamefont {Richerme}}, \bibinfo {author} {\bibfnamefont {B.}~\bibnamefont {Neyenhuis}}, \bibinfo {author} {\bibfnamefont {P.~W.}\ \bibnamefont {Hess}}, \bibinfo {author} {\bibfnamefont {P.}~\bibnamefont {Hauke}}, \bibinfo {author} {\bibfnamefont {M.}~\bibnamefont {Heyl}}, \bibinfo {author} {\bibfnamefont {D.~A.}\ \bibnamefont {Huse}},\ and\ \bibinfo {author} {\bibfnamefont {C.}~\bibnamefont {Monroe}},\ }\bibfield  {title} {\bibinfo {title} {Many-body localization in a quantum simulator with programmable random disorder},\ }\href {http://dx.doi.org/10.1038/nphys3783} {\bibfield  {journal} {\bibinfo  {journal} {Nat. Phys.}\ }\textbf {\bibinfo {volume} {12}},\ \bibinfo {pages} {907} (\bibinfo {year} {2016})}\BibitemShut {NoStop}%
\bibitem [{\citenamefont {Choi}\ \emph {et~al.}(2017)\citenamefont {Choi}, \citenamefont {Choi}, \citenamefont {Landig}, \citenamefont {Kucsko}, \citenamefont {Zhou}, \citenamefont {Isoya}, \citenamefont {Jelezko}, \citenamefont {Onoda}, \citenamefont {Sumiya}, \citenamefont {Khemani}, \citenamefont {von Keyserlingk}, \citenamefont {Yao}, \citenamefont {Demler},\ and\ \citenamefont {Lukin}}]{Choi2017}%
  \BibitemOpen
  \bibfield  {author} {\bibinfo {author} {\bibfnamefont {S.}~\bibnamefont {Choi}}, \bibinfo {author} {\bibfnamefont {J.}~\bibnamefont {Choi}}, \bibinfo {author} {\bibfnamefont {R.}~\bibnamefont {Landig}}, \bibinfo {author} {\bibfnamefont {G.}~\bibnamefont {Kucsko}}, \bibinfo {author} {\bibfnamefont {H.}~\bibnamefont {Zhou}}, \bibinfo {author} {\bibfnamefont {J.}~\bibnamefont {Isoya}}, \bibinfo {author} {\bibfnamefont {F.}~\bibnamefont {Jelezko}}, \bibinfo {author} {\bibfnamefont {S.}~\bibnamefont {Onoda}}, \bibinfo {author} {\bibfnamefont {H.}~\bibnamefont {Sumiya}}, \bibinfo {author} {\bibfnamefont {V.}~\bibnamefont {Khemani}}, \bibinfo {author} {\bibfnamefont {C.}~\bibnamefont {von Keyserlingk}}, \bibinfo {author} {\bibfnamefont {N.~Y.}\ \bibnamefont {Yao}}, \bibinfo {author} {\bibfnamefont {E.}~\bibnamefont {Demler}},\ and\ \bibinfo {author} {\bibfnamefont {M.~D.}\ \bibnamefont {Lukin}},\ }\bibfield  {title} {\bibinfo {title} {Observation of discrete time-crystalline order in a disordered dipolar many-body
  system},\ }\href {https://doi.org/10.1038/nature21426} {\bibfield  {journal} {\bibinfo  {journal} {Nature}\ }\textbf {\bibinfo {volume} {543}},\ \bibinfo {pages} {221} (\bibinfo {year} {2017})}\BibitemShut {NoStop}%
\bibitem [{\citenamefont {Roushan}\ \emph {et~al.}(2017)\citenamefont {Roushan}, \citenamefont {Neill}, \citenamefont {Tangpanitanon}, \citenamefont {Bastidas}, \citenamefont {Megrant}, \citenamefont {Barends}, \citenamefont {Chen}, \citenamefont {Chen}, \citenamefont {Chiaro}, \citenamefont {Dunsworth} \emph {et~al.}}]{Roushan2017}%
  \BibitemOpen
  \bibfield  {author} {\bibinfo {author} {\bibfnamefont {P.}~\bibnamefont {Roushan}}, \bibinfo {author} {\bibfnamefont {C.}~\bibnamefont {Neill}}, \bibinfo {author} {\bibfnamefont {J.}~\bibnamefont {Tangpanitanon}}, \bibinfo {author} {\bibfnamefont {V.~M.}\ \bibnamefont {Bastidas}}, \bibinfo {author} {\bibfnamefont {A.}~\bibnamefont {Megrant}}, \bibinfo {author} {\bibfnamefont {R.}~\bibnamefont {Barends}}, \bibinfo {author} {\bibfnamefont {Y.}~\bibnamefont {Chen}}, \bibinfo {author} {\bibfnamefont {Z.}~\bibnamefont {Chen}}, \bibinfo {author} {\bibfnamefont {B.}~\bibnamefont {Chiaro}}, \bibinfo {author} {\bibfnamefont {A.}~\bibnamefont {Dunsworth}}, \emph {et~al.},\ }\bibfield  {title} {\bibinfo {title} {Spectroscopic signatures of localization with interacting photons in superconducting qubits},\ }\href {https://doi.org/10.1126/science.aao1401} {\bibfield  {journal} {\bibinfo  {journal} {Science}\ }\textbf {\bibinfo {volume} {358}},\ \bibinfo {pages} {1175} (\bibinfo {year} {2017})}\BibitemShut {NoStop}%
\bibitem [{\citenamefont {Lukin}\ \emph {et~al.}(2019)\citenamefont {Lukin}, \citenamefont {Rispoli}, \citenamefont {Schittko}, \citenamefont {Tai}, \citenamefont {Kaufman}, \citenamefont {Choi}, \citenamefont {Khemani}, \citenamefont {L\'{e}onard},\ and\ \citenamefont {Greiner}}]{Lukin2018}%
  \BibitemOpen
  \bibfield  {author} {\bibinfo {author} {\bibfnamefont {A.}~\bibnamefont {Lukin}}, \bibinfo {author} {\bibfnamefont {M.}~\bibnamefont {Rispoli}}, \bibinfo {author} {\bibfnamefont {R.}~\bibnamefont {Schittko}}, \bibinfo {author} {\bibfnamefont {M.~E.}\ \bibnamefont {Tai}}, \bibinfo {author} {\bibfnamefont {A.~M.}\ \bibnamefont {Kaufman}}, \bibinfo {author} {\bibfnamefont {S.}~\bibnamefont {Choi}}, \bibinfo {author} {\bibfnamefont {V.}~\bibnamefont {Khemani}}, \bibinfo {author} {\bibfnamefont {J.}~\bibnamefont {L\'{e}onard}},\ and\ \bibinfo {author} {\bibfnamefont {M.}~\bibnamefont {Greiner}},\ }\bibfield  {title} {\bibinfo {title} {Probing entanglement in a many-body-localized system},\ }\href {https://doi.org/10.1126/science.aau0818} {\bibfield  {journal} {\bibinfo  {journal} {Science}\ }\textbf {\bibinfo {volume} {364}},\ \bibinfo {pages} {256} (\bibinfo {year} {2019})}\BibitemShut {NoStop}%
\bibitem [{\citenamefont {Imbrie}(2016)}]{imbrie2016many}%
  \BibitemOpen
  \bibfield  {author} {\bibinfo {author} {\bibfnamefont {J.~Z.}\ \bibnamefont {Imbrie}},\ }\bibfield  {title} {\bibinfo {title} {On many-body localization for quantum spin chains},\ }\href {https://doi.org/10.1007/s10955-016-1508-x} {\bibfield  {journal} {\bibinfo  {journal} {J. Stat. Phys.}\ }\textbf {\bibinfo {volume} {163}},\ \bibinfo {pages} {998} (\bibinfo {year} {2016})}\BibitemShut {NoStop}%
\bibitem [{\citenamefont {Šuntajs}\ \emph {et~al.}(2020)\citenamefont {Šuntajs}, \citenamefont {Bonča}, \citenamefont {Prosen},\ and\ \citenamefont {Vidmar}}]{untajs2020}%
  \BibitemOpen
  \bibfield  {author} {\bibinfo {author} {\bibfnamefont {J.}~\bibnamefont {Šuntajs}}, \bibinfo {author} {\bibfnamefont {J.}~\bibnamefont {Bonča}}, \bibinfo {author} {\bibfnamefont {T.}~\bibnamefont {Prosen}},\ and\ \bibinfo {author} {\bibfnamefont {L.}~\bibnamefont {Vidmar}},\ }\bibfield  {title} {\bibinfo {title} {Quantum chaos challenges many-body localization},\ }\href {https://doi.org/10.1103/physreve.102.062144} {\bibfield  {journal} {\bibinfo  {journal} {Phys. Rev. E}\ }\textbf {\bibinfo {volume} {102}},\ \bibinfo {pages} {062144} (\bibinfo {year} {2020})}\BibitemShut {NoStop}%
\bibitem [{\citenamefont {Sels}\ and\ \citenamefont {Polkovnikov}(2021)}]{Sels2021}%
  \BibitemOpen
  \bibfield  {author} {\bibinfo {author} {\bibfnamefont {D.}~\bibnamefont {Sels}}\ and\ \bibinfo {author} {\bibfnamefont {A.}~\bibnamefont {Polkovnikov}},\ }\bibfield  {title} {\bibinfo {title} {Dynamical obstruction to localization in a disordered spin chain},\ }\href {https://doi.org/10.1103/physreve.104.054105} {\bibfield  {journal} {\bibinfo  {journal} {Phys. Rev. E}\ }\textbf {\bibinfo {volume} {104}},\ \bibinfo {pages} {054105} (\bibinfo {year} {2021})}\BibitemShut {NoStop}%
\bibitem [{\citenamefont {Peacock}\ and\ \citenamefont {Sels}(2023)}]{Peacock2023}%
  \BibitemOpen
  \bibfield  {author} {\bibinfo {author} {\bibfnamefont {J.~C.}\ \bibnamefont {Peacock}}\ and\ \bibinfo {author} {\bibfnamefont {D.}~\bibnamefont {Sels}},\ }\bibfield  {title} {\bibinfo {title} {Many-body delocalization from embedded thermal inclusion},\ }\href {https://doi.org/10.1103/physrevb.108.l020201} {\bibfield  {journal} {\bibinfo  {journal} {Phys. Rev. B}\ }\textbf {\bibinfo {volume} {108}},\ \bibinfo {pages} {L020201} (\bibinfo {year} {2023})}\BibitemShut {NoStop}%
\bibitem [{\citenamefont {Roeck}\ and\ \citenamefont {Imbrie}(2017)}]{deRoeck2017Stability}%
  \BibitemOpen
  \bibfield  {author} {\bibinfo {author} {\bibfnamefont {W.~D.}\ \bibnamefont {Roeck}}\ and\ \bibinfo {author} {\bibfnamefont {J.~Z.}\ \bibnamefont {Imbrie}},\ }\bibfield  {title} {\bibinfo {title} {Many-body localization: stability and instability},\ }\href {https://doi.org/10.1098/rsta.2016.0422} {\bibfield  {journal} {\bibinfo  {journal} {Phil. Trans. R. Soc. A}\ }\textbf {\bibinfo {volume} {375}},\ \bibinfo {pages} {20160422} (\bibinfo {year} {2017})}\BibitemShut {NoStop}%
\bibitem [{\citenamefont {Gopalakrishnan}\ and\ \citenamefont {Huse}(2019)}]{Gopalakrishnan2019}%
  \BibitemOpen
  \bibfield  {author} {\bibinfo {author} {\bibfnamefont {S.}~\bibnamefont {Gopalakrishnan}}\ and\ \bibinfo {author} {\bibfnamefont {D.~A.}\ \bibnamefont {Huse}},\ }\bibfield  {title} {\bibinfo {title} {Instability of many-body localized systems as a phase transition in a nonstandard thermodynamic limit},\ }\href {https://doi.org/10.1103/physrevb.99.134305} {\bibfield  {journal} {\bibinfo  {journal} {Phys. Rev. B}\ }\textbf {\bibinfo {volume} {99}},\ \bibinfo {pages} {134305} (\bibinfo {year} {2019})}\BibitemShut {NoStop}%
\bibitem [{\citenamefont {Choi}\ \emph {et~al.}(2016)\citenamefont {Choi}, \citenamefont {Hild}, \citenamefont {Zeiher}, \citenamefont {Schau{\ss}}, \citenamefont {Rubio-Abadal}, \citenamefont {Yefsah}, \citenamefont {Khemani}, \citenamefont {Huse}, \citenamefont {Bloch},\ and\ \citenamefont {Gross}}]{Choi1547}%
  \BibitemOpen
  \bibfield  {author} {\bibinfo {author} {\bibfnamefont {J.-y.}\ \bibnamefont {Choi}}, \bibinfo {author} {\bibfnamefont {S.}~\bibnamefont {Hild}}, \bibinfo {author} {\bibfnamefont {J.}~\bibnamefont {Zeiher}}, \bibinfo {author} {\bibfnamefont {P.}~\bibnamefont {Schau{\ss}}}, \bibinfo {author} {\bibfnamefont {A.}~\bibnamefont {Rubio-Abadal}}, \bibinfo {author} {\bibfnamefont {T.}~\bibnamefont {Yefsah}}, \bibinfo {author} {\bibfnamefont {V.}~\bibnamefont {Khemani}}, \bibinfo {author} {\bibfnamefont {D.~A.}\ \bibnamefont {Huse}}, \bibinfo {author} {\bibfnamefont {I.}~\bibnamefont {Bloch}},\ and\ \bibinfo {author} {\bibfnamefont {C.}~\bibnamefont {Gross}},\ }\bibfield  {title} {\bibinfo {title} {Exploring the many-body localization transition in two dimensions},\ }\href {https://doi.org/10.1126/science.aaf8834} {\bibfield  {journal} {\bibinfo  {journal} {Science}\ }\textbf {\bibinfo {volume} {352}},\ \bibinfo {pages} {1547} (\bibinfo {year} {2016})}\BibitemShut {NoStop}%
\bibitem [{\citenamefont {Rubio-Abadal}\ \emph {et~al.}(2019)\citenamefont {Rubio-Abadal}, \citenamefont {Choi}, \citenamefont {Zeiher}, \citenamefont {Hollerith}, \citenamefont {Rui}, \citenamefont {Bloch},\ and\ \citenamefont {Gross}}]{cleanDirty2DMBL}%
  \BibitemOpen
  \bibfield  {author} {\bibinfo {author} {\bibfnamefont {A.}~\bibnamefont {Rubio-Abadal}}, \bibinfo {author} {\bibfnamefont {J.-y.}\ \bibnamefont {Choi}}, \bibinfo {author} {\bibfnamefont {J.}~\bibnamefont {Zeiher}}, \bibinfo {author} {\bibfnamefont {S.}~\bibnamefont {Hollerith}}, \bibinfo {author} {\bibfnamefont {J.}~\bibnamefont {Rui}}, \bibinfo {author} {\bibfnamefont {I.}~\bibnamefont {Bloch}},\ and\ \bibinfo {author} {\bibfnamefont {C.}~\bibnamefont {Gross}},\ }\bibfield  {title} {\bibinfo {title} {{Many-Body Delocalization in the Presence of a Quantum Bath}},\ }\href {https://doi.org/10.1103/physrevx.9.041014} {\bibfield  {journal} {\bibinfo  {journal} {Phys. Rev. X}\ }\textbf {\bibinfo {volume} {9}},\ \bibinfo {pages} {041014} (\bibinfo {year} {2019})}\BibitemShut {NoStop}%
\bibitem [{\citenamefont {Prabhu}\ and\ \citenamefont {Mueller}(2021)}]{prabhu2021bathmediateddecaydensity}%
  \BibitemOpen
  \bibfield  {author} {\bibinfo {author} {\bibfnamefont {S.}~\bibnamefont {Prabhu}}\ and\ \bibinfo {author} {\bibfnamefont {E.~J.}\ \bibnamefont {Mueller}},\ }\href {https://arxiv.org/abs/2104.12648} {\bibinfo {title} {Bath mediated decay of density waves in a disordered bose lattice gas}} (\bibinfo {year} {2021}),\ \Eprint {https://arxiv.org/abs/2104.12648} {arXiv:2104.12648 [cond-mat.quant-gas]} \BibitemShut {NoStop}%
\bibitem [{\citenamefont {Yan}\ \emph {et~al.}(2017)\citenamefont {Yan}, \citenamefont {Hui}, \citenamefont {Rigol},\ and\ \citenamefont {Scarola}}]{glassyMBL}%
  \BibitemOpen
  \bibfield  {author} {\bibinfo {author} {\bibfnamefont {M.}~\bibnamefont {Yan}}, \bibinfo {author} {\bibfnamefont {H.-Y.}\ \bibnamefont {Hui}}, \bibinfo {author} {\bibfnamefont {M.}~\bibnamefont {Rigol}},\ and\ \bibinfo {author} {\bibfnamefont {V.~W.}\ \bibnamefont {Scarola}},\ }\bibfield  {title} {\bibinfo {title} {Equilibration dynamics of strongly interacting bosons in 2d lattices with disorder},\ }\href {https://doi.org/10.1103/PhysRevLett.119.073002} {\bibfield  {journal} {\bibinfo  {journal} {Phys. Rev. Lett.}\ }\textbf {\bibinfo {volume} {119}},\ \bibinfo {pages} {073002} (\bibinfo {year} {2017})}\BibitemShut {NoStop}%
\bibitem [{\citenamefont {Pollmann}\ \emph {et~al.}(2016)\citenamefont {Pollmann}, \citenamefont {Khemani}, \citenamefont {Cirac},\ and\ \citenamefont {Sondhi}}]{Pollmann2016TNS}%
  \BibitemOpen
  \bibfield  {author} {\bibinfo {author} {\bibfnamefont {F.}~\bibnamefont {Pollmann}}, \bibinfo {author} {\bibfnamefont {V.}~\bibnamefont {Khemani}}, \bibinfo {author} {\bibfnamefont {J.~I.}\ \bibnamefont {Cirac}},\ and\ \bibinfo {author} {\bibfnamefont {S.~L.}\ \bibnamefont {Sondhi}},\ }\bibfield  {title} {\bibinfo {title} {{Efficient variational diagonalization of fully many-body localized Hamiltonians}},\ }\href {https://doi.org/10.1103/PhysRevB.94.041116} {\bibfield  {journal} {\bibinfo  {journal} {Phys. Rev. B}\ }\textbf {\bibinfo {volume} {94}},\ \bibinfo {pages} {041116(R)} (\bibinfo {year} {2016})}\BibitemShut {NoStop}%
\bibitem [{\citenamefont {Wahl}\ \emph {et~al.}(2017)\citenamefont {Wahl}, \citenamefont {Pal},\ and\ \citenamefont {Simon}}]{Wahl2017PRX}%
  \BibitemOpen
  \bibfield  {author} {\bibinfo {author} {\bibfnamefont {T.~B.}\ \bibnamefont {Wahl}}, \bibinfo {author} {\bibfnamefont {A.}~\bibnamefont {Pal}},\ and\ \bibinfo {author} {\bibfnamefont {S.~H.}\ \bibnamefont {Simon}},\ }\bibfield  {title} {\bibinfo {title} {{Efficient Representation of Fully Many-Body Localized Systems Using Tensor Networks}},\ }\href {https://doi.org/10.1103/PhysRevX.7.021018} {\bibfield  {journal} {\bibinfo  {journal} {Phys. Rev. X}\ }\textbf {\bibinfo {volume} {7}},\ \bibinfo {pages} {021018} (\bibinfo {year} {2017})}\BibitemShut {NoStop}%
\bibitem [{\citenamefont {Wahl}\ \emph {et~al.}(2022)\citenamefont {Wahl}, \citenamefont {Venn},\ and\ \citenamefont {B{\'{e}}ri}}]{Wahl2022}%
  \BibitemOpen
  \bibfield  {author} {\bibinfo {author} {\bibfnamefont {T.~B.}\ \bibnamefont {Wahl}}, \bibinfo {author} {\bibfnamefont {F.}~\bibnamefont {Venn}},\ and\ \bibinfo {author} {\bibfnamefont {B.}~\bibnamefont {B{\'{e}}ri}},\ }\bibfield  {title} {\bibinfo {title} {Local integrals of motion detection of localization-protected topological order},\ }\href {https://doi.org/10.1103/physrevb.105.144205} {\bibfield  {journal} {\bibinfo  {journal} {Phys. Rev. B}\ }\textbf {\bibinfo {volume} {105}},\ \bibinfo {pages} {144205} (\bibinfo {year} {2022})}\BibitemShut {NoStop}%
\bibitem [{\citenamefont {Wahl}\ \emph {et~al.}(2019)\citenamefont {Wahl}, \citenamefont {Pal},\ and\ \citenamefont {Simon}}]{2DMBL}%
  \BibitemOpen
  \bibfield  {author} {\bibinfo {author} {\bibfnamefont {T.~B.}\ \bibnamefont {Wahl}}, \bibinfo {author} {\bibfnamefont {A.}~\bibnamefont {Pal}},\ and\ \bibinfo {author} {\bibfnamefont {S.~H.}\ \bibnamefont {Simon}},\ }\bibfield  {title} {\bibinfo {title} {{Signatures of the Many-body Localized Regime in Two Dimensions}},\ }\href {https://doi.org/10.1038/s41567-018-0339-x} {\bibfield  {journal} {\bibinfo  {journal} {Nat. Phys.}\ }\textbf {\bibinfo {volume} {15}},\ \bibinfo {pages} {164} (\bibinfo {year} {2019})}\BibitemShut {NoStop}%
\bibitem [{\citenamefont {Li}\ \emph {et~al.}(2024)\citenamefont {Li}, \citenamefont {Chan},\ and\ \citenamefont {Wahl}}]{2dMBLfermion}%
  \BibitemOpen
  \bibfield  {author} {\bibinfo {author} {\bibfnamefont {J.}~\bibnamefont {Li}}, \bibinfo {author} {\bibfnamefont {A.}~\bibnamefont {Chan}},\ and\ \bibinfo {author} {\bibfnamefont {T.~B.}\ \bibnamefont {Wahl}},\ }\bibfield  {title} {\bibinfo {title} {Quantum circuits reproduce the experimental two-dimensional many-body localization transition point},\ }\href {https://doi.org/10.1103/PhysRevB.109.L140202} {\bibfield  {journal} {\bibinfo  {journal} {Phys. Rev. B}\ }\textbf {\bibinfo {volume} {109}},\ \bibinfo {pages} {L140202} (\bibinfo {year} {2024})}\BibitemShut {NoStop}%
\bibitem [{\citenamefont {Venn}\ \emph {et~al.}()\citenamefont {Venn}, \citenamefont {Wahl},\ and\ \citenamefont {B\'{e}ri}}]{Venn2023}%
  \BibitemOpen
  \bibfield  {author} {\bibinfo {author} {\bibfnamefont {F.}~\bibnamefont {Venn}}, \bibinfo {author} {\bibfnamefont {T.~B.}\ \bibnamefont {Wahl}},\ and\ \bibinfo {author} {\bibfnamefont {B.}~\bibnamefont {B\'{e}ri}},\ }\bibfield  {title} {\bibinfo {title} {Many-body-localization protection of eigenstate topological order in two dimensions},\ }\href {https://doi.org/10.48550/arXiv.2212.09775} {\bibinfo  {journal} {arXiv:2212.09775, to appear in Phys. Rev. B}\ }\BibitemShut {NoStop}%
\bibitem [{\citenamefont {Wahl}(2018)}]{Thorsten}%
  \BibitemOpen
\bibfield  {journal} {  }\bibfield  {author} {\bibinfo {author} {\bibfnamefont {T.~B.}\ \bibnamefont {Wahl}},\ }\bibfield  {title} {\bibinfo {title} {Tensor networks demonstrate the robustness of localization and symmetry-protected topological phases},\ }\href {https://doi.org/10.1103/physrevb.98.054204} {\bibfield  {journal} {\bibinfo  {journal} {Phys. Rev. B}\ }\textbf {\bibinfo {volume} {98}},\ \bibinfo {pages} {054204} (\bibinfo {year} {2018})}\BibitemShut {NoStop}%
\bibitem [{\citenamefont {Chan}\ and\ \citenamefont {Wahl}(2020)}]{1DSPTMBL}%
  \BibitemOpen
  \bibfield  {author} {\bibinfo {author} {\bibfnamefont {A.}~\bibnamefont {Chan}}\ and\ \bibinfo {author} {\bibfnamefont {T.~B.}\ \bibnamefont {Wahl}},\ }\bibfield  {title} {\bibinfo {title} {Classification of symmetry-protected topological many-body localized phases in one dimension},\ }\href {https://doi.org/10.1088/1361-648x/ab7f01} {\bibfield  {journal} {\bibinfo  {journal} {J. Phys.: Cond. Mat.}\ }\textbf {\bibinfo {volume} {32}},\ \bibinfo {pages} {305601} (\bibinfo {year} {2020})}\BibitemShut {NoStop}%
\bibitem [{\citenamefont {Li}\ \emph {et~al.}(2020)\citenamefont {Li}, \citenamefont {Chan},\ and\ \citenamefont {Wahl}}]{2DSPTMBL}%
  \BibitemOpen
  \bibfield  {author} {\bibinfo {author} {\bibfnamefont {J.}~\bibnamefont {Li}}, \bibinfo {author} {\bibfnamefont {A.}~\bibnamefont {Chan}},\ and\ \bibinfo {author} {\bibfnamefont {T.~B.}\ \bibnamefont {Wahl}},\ }\bibfield  {title} {\bibinfo {title} {Classification of symmetry-protected topological phases in two-dimensional many-body localized systems},\ }\href {https://doi.org/10.1103/physrevb.102.014205} {\bibfield  {journal} {\bibinfo  {journal} {Phys. Rev. B}\ }\textbf {\bibinfo {volume} {102}},\ \bibinfo {pages} {014205} (\bibinfo {year} {2020})}\BibitemShut {NoStop}%
\bibitem [{\citenamefont {Haegeman}\ \emph {et~al.}(2016)\citenamefont {Haegeman}, \citenamefont {Lubich}, \citenamefont {Oseledets}, \citenamefont {Vandereycken},\ and\ \citenamefont {Verstraete}}]{tdvp}%
  \BibitemOpen
  \bibfield  {author} {\bibinfo {author} {\bibfnamefont {J.}~\bibnamefont {Haegeman}}, \bibinfo {author} {\bibfnamefont {C.}~\bibnamefont {Lubich}}, \bibinfo {author} {\bibfnamefont {I.}~\bibnamefont {Oseledets}}, \bibinfo {author} {\bibfnamefont {B.}~\bibnamefont {Vandereycken}},\ and\ \bibinfo {author} {\bibfnamefont {F.}~\bibnamefont {Verstraete}},\ }\bibfield  {title} {\bibinfo {title} {Unifying time evolution and optimization with matrix product states},\ }\href {https://doi.org/10.1103/PhysRevB.94.165116} {\bibfield  {journal} {\bibinfo  {journal} {Phys. Rev. B}\ }\textbf {\bibinfo {volume} {94}},\ \bibinfo {pages} {165116} (\bibinfo {year} {2016})}\BibitemShut {NoStop}%
\bibitem [{\citenamefont {Doggen}\ \emph {et~al.}(2020)\citenamefont {Doggen}, \citenamefont {Gornyi}, \citenamefont {Mirlin},\ and\ \citenamefont {Polyakov}}]{Elmer}%
  \BibitemOpen
  \bibfield  {author} {\bibinfo {author} {\bibfnamefont {E.~V.~H.}\ \bibnamefont {Doggen}}, \bibinfo {author} {\bibfnamefont {I.~V.}\ \bibnamefont {Gornyi}}, \bibinfo {author} {\bibfnamefont {A.~D.}\ \bibnamefont {Mirlin}},\ and\ \bibinfo {author} {\bibfnamefont {D.~G.}\ \bibnamefont {Polyakov}},\ }\bibfield  {title} {\bibinfo {title} {Slow many-body delocalization beyond one dimension},\ }\href {https://doi.org/10.1103/PhysRevLett.125.155701} {\bibfield  {journal} {\bibinfo  {journal} {Phys. Rev. Lett.}\ }\textbf {\bibinfo {volume} {125}},\ \bibinfo {pages} {155701} (\bibinfo {year} {2020})}\BibitemShut {NoStop}%
\bibitem [{\citenamefont {Crowley}\ and\ \citenamefont {Chandran}(2022)}]{Crowley2022}%
  \BibitemOpen
  \bibfield  {author} {\bibinfo {author} {\bibfnamefont {P.}~\bibnamefont {Crowley}}\ and\ \bibinfo {author} {\bibfnamefont {A.}~\bibnamefont {Chandran}},\ }\bibfield  {title} {\bibinfo {title} {A constructive theory of the numerically accessible many-body localized to thermal crossover},\ }\href {https://doi.org/10.21468/scipostphys.12.6.201} {\bibfield  {journal} {\bibinfo  {journal} {SciPost Physics}\ }\textbf {\bibinfo {volume} {12}},\ \bibinfo {pages} {201} (\bibinfo {year} {2022})}\BibitemShut {NoStop}%
\bibitem [{\citenamefont {Serbyn}\ \emph {et~al.}(2013)\citenamefont {Serbyn}, \citenamefont {Papi{\'{c}}},\ and\ \citenamefont {Abanin}}]{serbyn2013local}%
  \BibitemOpen
  \bibfield  {author} {\bibinfo {author} {\bibfnamefont {M.}~\bibnamefont {Serbyn}}, \bibinfo {author} {\bibfnamefont {Z.}~\bibnamefont {Papi{\'{c}}}},\ and\ \bibinfo {author} {\bibfnamefont {D.~A.}\ \bibnamefont {Abanin}},\ }\bibfield  {title} {\bibinfo {title} {{Local Conservation Laws and the Structure of the Many-Body Localized States}},\ }\href {https://doi.org/10.1103/physrevlett.111.127201} {\bibfield  {journal} {\bibinfo  {journal} {Phys. Rev. Lett.}\ }\textbf {\bibinfo {volume} {111}},\ \bibinfo {pages} {127201} (\bibinfo {year} {2013})}\BibitemShut {NoStop}%
\bibitem [{\citenamefont {Huse}\ \emph {et~al.}(2014)\citenamefont {Huse}, \citenamefont {Nandkishore},\ and\ \citenamefont {Oganesyan}}]{Huse_MBL_phenom_14}%
  \BibitemOpen
  \bibfield  {author} {\bibinfo {author} {\bibfnamefont {D.~A.}\ \bibnamefont {Huse}}, \bibinfo {author} {\bibfnamefont {R.}~\bibnamefont {Nandkishore}},\ and\ \bibinfo {author} {\bibfnamefont {V.}~\bibnamefont {Oganesyan}},\ }\bibfield  {title} {\bibinfo {title} {Phenomenology of fully many-body-localized systems},\ }\href {https://doi.org/10.1103/PhysRevB.90.174202} {\bibfield  {journal} {\bibinfo  {journal} {Phys. Rev. B}\ }\textbf {\bibinfo {volume} {90}},\ \bibinfo {pages} {174202} (\bibinfo {year} {2014})}\BibitemShut {NoStop}%
\bibitem [{\citenamefont {Chandran}\ \emph {et~al.}(2015)\citenamefont {Chandran}, \citenamefont {Kim}, \citenamefont {Vidal},\ and\ \citenamefont {Abanin}}]{chandran2015constructing}%
  \BibitemOpen
  \bibfield  {author} {\bibinfo {author} {\bibfnamefont {A.}~\bibnamefont {Chandran}}, \bibinfo {author} {\bibfnamefont {I.~H.}\ \bibnamefont {Kim}}, \bibinfo {author} {\bibfnamefont {G.}~\bibnamefont {Vidal}},\ and\ \bibinfo {author} {\bibfnamefont {D.~A.}\ \bibnamefont {Abanin}},\ }\bibfield  {title} {\bibinfo {title} {Constructing local integrals of motion in the many-body localized phase},\ }\href {https://doi.org/10.1103/PhysRevB.91.085425} {\bibfield  {journal} {\bibinfo  {journal} {Phys. Rev. B}\ }\textbf {\bibinfo {volume} {91}},\ \bibinfo {pages} {085425} (\bibinfo {year} {2015})}\BibitemShut {NoStop}%
\bibitem [{\citenamefont {Ros}\ \emph {et~al.}(2015)\citenamefont {Ros}, \citenamefont {M\"{u}ller},\ and\ \citenamefont {Scardicchio}}]{ros2015integrals}%
  \BibitemOpen
  \bibfield  {author} {\bibinfo {author} {\bibfnamefont {V.}~\bibnamefont {Ros}}, \bibinfo {author} {\bibfnamefont {M.}~\bibnamefont {M\"{u}ller}},\ and\ \bibinfo {author} {\bibfnamefont {A.}~\bibnamefont {Scardicchio}},\ }\bibfield  {title} {\bibinfo {title} {Integrals of motion in the many-body localized phase},\ }\href {https://doi.org/10.1016/j.nuclphysb.2014.12.014} {\bibfield  {journal} {\bibinfo  {journal} {Nucl. Phys. B}\ }\textbf {\bibinfo {volume} {891}},\ \bibinfo {pages} {420} (\bibinfo {year} {2015})}\BibitemShut {NoStop}%
\bibitem [{\citenamefont {Inglis}\ and\ \citenamefont {Pollet}(2016)}]{Inglis_PRL2016}%
  \BibitemOpen
  \bibfield  {author} {\bibinfo {author} {\bibfnamefont {S.}~\bibnamefont {Inglis}}\ and\ \bibinfo {author} {\bibfnamefont {L.}~\bibnamefont {Pollet}},\ }\bibfield  {title} {\bibinfo {title} {{Accessing Many-Body Localized States through the Generalized Gibbs Ensemble}},\ }\href {https://doi.org/10.1103/PhysRevLett.117.120402} {\bibfield  {journal} {\bibinfo  {journal} {Phys. Rev. Lett.}\ }\textbf {\bibinfo {volume} {117}},\ \bibinfo {pages} {120402} (\bibinfo {year} {2016})}\BibitemShut {NoStop}%
\bibitem [{\citenamefont {Rademaker}\ and\ \citenamefont {Ortu\~no}(2016)}]{Rademaker2016LIOM}%
  \BibitemOpen
  \bibfield  {author} {\bibinfo {author} {\bibfnamefont {L.}~\bibnamefont {Rademaker}}\ and\ \bibinfo {author} {\bibfnamefont {M.}~\bibnamefont {Ortu\~no}},\ }\bibfield  {title} {\bibinfo {title} {{Explicit Local Integrals of Motion for the Many-Body Localized State}},\ }\href {https://doi.org/10.1103/PhysRevLett.116.010404} {\bibfield  {journal} {\bibinfo  {journal} {Phys. Rev. Lett.}\ }\textbf {\bibinfo {volume} {116}},\ \bibinfo {pages} {010404} (\bibinfo {year} {2016})}\BibitemShut {NoStop}%
\bibitem [{\citenamefont {Monthus}(2016)}]{Monthus2016}%
  \BibitemOpen
  \bibfield  {author} {\bibinfo {author} {\bibfnamefont {C.}~\bibnamefont {Monthus}},\ }\bibfield  {title} {\bibinfo {title} {Many-body localization: construction of the emergent local conserved operators via block real-space renormalization},\ }\href {https://doi.org/10.1088/1742-5468/2016/03/033101} {\bibfield  {journal} {\bibinfo  {journal} {J. Stat. Mech.}\ }\textbf {\bibinfo {volume} {2016}},\ \bibinfo {pages} {033101} (\bibinfo {year} {2016})}\BibitemShut {NoStop}%
\bibitem [{\citenamefont {Goihl}\ \emph {et~al.}(2018)\citenamefont {Goihl}, \citenamefont {Gluza}, \citenamefont {Krumnow},\ and\ \citenamefont {Eisert}}]{Goihl2018}%
  \BibitemOpen
  \bibfield  {author} {\bibinfo {author} {\bibfnamefont {M.}~\bibnamefont {Goihl}}, \bibinfo {author} {\bibfnamefont {M.}~\bibnamefont {Gluza}}, \bibinfo {author} {\bibfnamefont {C.}~\bibnamefont {Krumnow}},\ and\ \bibinfo {author} {\bibfnamefont {J.}~\bibnamefont {Eisert}},\ }\bibfield  {title} {\bibinfo {title} {Construction of exact constants of motion and effective models for many-body localized systems},\ }\href {https://doi.org/10.1103/PhysRevB.97.134202} {\bibfield  {journal} {\bibinfo  {journal} {Phys. Rev. B}\ }\textbf {\bibinfo {volume} {97}},\ \bibinfo {pages} {134202} (\bibinfo {year} {2018})}\BibitemShut {NoStop}%
\bibitem [{\citenamefont {Kulshreshtha}\ \emph {et~al.}(2018)\citenamefont {Kulshreshtha}, \citenamefont {Pal}, \citenamefont {Wahl},\ and\ \citenamefont {Simon}}]{Abi2017}%
  \BibitemOpen
  \bibfield  {author} {\bibinfo {author} {\bibfnamefont {A.~K.}\ \bibnamefont {Kulshreshtha}}, \bibinfo {author} {\bibfnamefont {A.}~\bibnamefont {Pal}}, \bibinfo {author} {\bibfnamefont {T.~B.}\ \bibnamefont {Wahl}},\ and\ \bibinfo {author} {\bibfnamefont {S.~H.}\ \bibnamefont {Simon}},\ }\bibfield  {title} {\bibinfo {title} {Behavior of l-bits near the many-body localization transition},\ }\href {https://doi.org/10.1103/physrevb.98.184201} {\bibfield  {journal} {\bibinfo  {journal} {Phys. Rev. B}\ }\textbf {\bibinfo {volume} {98}},\ \bibinfo {pages} {184201} (\bibinfo {year} {2018})}\BibitemShut {NoStop}%
\bibitem [{\citenamefont {Chandran}\ \emph {et~al.}(2016)\citenamefont {Chandran}, \citenamefont {Pal}, \citenamefont {Laumann},\ and\ \citenamefont {Scardicchio}}]{chandran2016higherD}%
  \BibitemOpen
  \bibfield  {author} {\bibinfo {author} {\bibfnamefont {A.}~\bibnamefont {Chandran}}, \bibinfo {author} {\bibfnamefont {A.}~\bibnamefont {Pal}}, \bibinfo {author} {\bibfnamefont {C.~R.}\ \bibnamefont {Laumann}},\ and\ \bibinfo {author} {\bibfnamefont {A.}~\bibnamefont {Scardicchio}},\ }\bibfield  {title} {\bibinfo {title} {Many-body localization beyond eigenstates in all dimensions},\ }\href {https://doi.org/10.1103/PhysRevB.94.144203} {\bibfield  {journal} {\bibinfo  {journal} {Phys. Rev. B}\ }\textbf {\bibinfo {volume} {94}},\ \bibinfo {pages} {144203} (\bibinfo {year} {2016})}\BibitemShut {NoStop}%
\bibitem [{\citenamefont {Luitz}\ \emph {et~al.}(2015)\citenamefont {Luitz}, \citenamefont {Laflorencie},\ and\ \citenamefont {Alet}}]{Luitz2015}%
  \BibitemOpen
  \bibfield  {author} {\bibinfo {author} {\bibfnamefont {D.~J.}\ \bibnamefont {Luitz}}, \bibinfo {author} {\bibfnamefont {N.}~\bibnamefont {Laflorencie}},\ and\ \bibinfo {author} {\bibfnamefont {F.}~\bibnamefont {Alet}},\ }\bibfield  {title} {\bibinfo {title} {{Many-body localization edge in the random-field Heisenberg chain}},\ }\href {https://doi.org/10.1103/physrevb.91.081103} {\bibfield  {journal} {\bibinfo  {journal} {Phys. Rev. B}\ }\textbf {\bibinfo {volume} {91}},\ \bibinfo {pages} {081103} (\bibinfo {year} {2015})}\BibitemShut {NoStop}%
\bibitem [{\citenamefont {Kj{\"a}ll}\ \emph {et~al.}(2014)\citenamefont {Kj{\"a}ll}, \citenamefont {Bardarson},\ and\ \citenamefont {Pollmann}}]{kjall2014many}%
  \BibitemOpen
  \bibfield  {author} {\bibinfo {author} {\bibfnamefont {J.~A.}\ \bibnamefont {Kj{\"a}ll}}, \bibinfo {author} {\bibfnamefont {J.~H.}\ \bibnamefont {Bardarson}},\ and\ \bibinfo {author} {\bibfnamefont {F.}~\bibnamefont {Pollmann}},\ }\bibfield  {title} {\bibinfo {title} {{Many-body localization in a disordered quantum Ising chain}},\ }\href {https://doi.org/10.1103/PhysRevLett.113.107204} {\bibfield  {journal} {\bibinfo  {journal} {Phys. Rev. Lett.}\ }\textbf {\bibinfo {volume} {113}},\ \bibinfo {pages} {107204} (\bibinfo {year} {2014})}\BibitemShut {NoStop}%
\bibitem [{\citenamefont {Hauschild}\ \emph {et~al.}(2024)\citenamefont {Hauschild}, \citenamefont {Unfried}, \citenamefont {Anand}, \citenamefont {Andrews}, \citenamefont {Bintz}, \citenamefont {Borla}, \citenamefont {Divic}, \citenamefont {Drescher}, \citenamefont {Geiger}, \citenamefont {Hefel}, \citenamefont {Hémery}, \citenamefont {Kadow}, \citenamefont {Kemp}, \citenamefont {Kirchner}, \citenamefont {Liu}, \citenamefont {Möller}, \citenamefont {Parker}, \citenamefont {Rader}, \citenamefont {Romen}, \citenamefont {Scalet}, \citenamefont {Schoonderwoerd}, \citenamefont {Schulz}, \citenamefont {Soejima}, \citenamefont {Thoma}, \citenamefont {Wu}, \citenamefont {Zechmann}, \citenamefont {Zweng}, \citenamefont {Mong}, \citenamefont {Zaletel},\ and\ \citenamefont {Pollmann}}]{tenpy}%
  \BibitemOpen
  \bibfield  {author} {\bibinfo {author} {\bibfnamefont {J.}~\bibnamefont {Hauschild}}, \bibinfo {author} {\bibfnamefont {J.}~\bibnamefont {Unfried}}, \bibinfo {author} {\bibfnamefont {S.}~\bibnamefont {Anand}}, \bibinfo {author} {\bibfnamefont {B.}~\bibnamefont {Andrews}}, \bibinfo {author} {\bibfnamefont {M.}~\bibnamefont {Bintz}}, \bibinfo {author} {\bibfnamefont {U.}~\bibnamefont {Borla}}, \bibinfo {author} {\bibfnamefont {S.}~\bibnamefont {Divic}}, \bibinfo {author} {\bibfnamefont {M.}~\bibnamefont {Drescher}}, \bibinfo {author} {\bibfnamefont {J.}~\bibnamefont {Geiger}}, \bibinfo {author} {\bibfnamefont {M.}~\bibnamefont {Hefel}}, \bibinfo {author} {\bibfnamefont {K.}~\bibnamefont {Hémery}}, \bibinfo {author} {\bibfnamefont {W.}~\bibnamefont {Kadow}}, \bibinfo {author} {\bibfnamefont {J.}~\bibnamefont {Kemp}}, \bibinfo {author} {\bibfnamefont {N.}~\bibnamefont {Kirchner}}, \bibinfo {author} {\bibfnamefont {V.~S.}\ \bibnamefont {Liu}}, \bibinfo {author} {\bibfnamefont {G.}~\bibnamefont {Möller}},
  \bibinfo {author} {\bibfnamefont {D.}~\bibnamefont {Parker}}, \bibinfo {author} {\bibfnamefont {M.}~\bibnamefont {Rader}}, \bibinfo {author} {\bibfnamefont {A.}~\bibnamefont {Romen}}, \bibinfo {author} {\bibfnamefont {S.}~\bibnamefont {Scalet}}, \bibinfo {author} {\bibfnamefont {L.}~\bibnamefont {Schoonderwoerd}}, \bibinfo {author} {\bibfnamefont {M.}~\bibnamefont {Schulz}}, \bibinfo {author} {\bibfnamefont {T.}~\bibnamefont {Soejima}}, \bibinfo {author} {\bibfnamefont {P.}~\bibnamefont {Thoma}}, \bibinfo {author} {\bibfnamefont {Y.}~\bibnamefont {Wu}}, \bibinfo {author} {\bibfnamefont {P.}~\bibnamefont {Zechmann}}, \bibinfo {author} {\bibfnamefont {L.}~\bibnamefont {Zweng}}, \bibinfo {author} {\bibfnamefont {R.~S.~K.}\ \bibnamefont {Mong}}, \bibinfo {author} {\bibfnamefont {M.~P.}\ \bibnamefont {Zaletel}},\ and\ \bibinfo {author} {\bibfnamefont {F.}~\bibnamefont {Pollmann}},\ }\bibfield  {title} {\bibinfo {title} {{Tensor network Python (TeNPy) version 1}},\ }\href
  {https://doi.org/10.21468/SciPostPhysCodeb.41} {\bibfield  {journal} {\bibinfo  {journal} {SciPost Phys. Codebases}\ ,\ \bibinfo {pages} {41}} (\bibinfo {year} {2024})}\BibitemShut {NoStop}%
\bibitem [{\citenamefont {Hundertmark}(2008)}]{Hundertmark}%
  \BibitemOpen
  \bibfield  {author} {\bibinfo {author} {\bibfnamefont {D.}~\bibnamefont {Hundertmark}},\ }\href {https://doi.org/10.1093/acprof:oso/9780199239252.003.0009} {\emph {\bibinfo {title} {{A short introduction to Anderson localization}}}}\ (\bibinfo  {publisher} {Analysis and stochastics of growth processes and interface models},\ \bibinfo {year} {2008})\ pp.\ \bibinfo {pages} {194--218}\BibitemShut {NoStop}%
\bibitem [{\citenamefont {Pietracaprina}\ \emph {et~al.}(2016)\citenamefont {Pietracaprina}, \citenamefont {Ros},\ and\ \citenamefont {Scardicchio}}]{Pietracaprina2016}%
  \BibitemOpen
  \bibfield  {author} {\bibinfo {author} {\bibfnamefont {F.}~\bibnamefont {Pietracaprina}}, \bibinfo {author} {\bibfnamefont {V.}~\bibnamefont {Ros}},\ and\ \bibinfo {author} {\bibfnamefont {A.}~\bibnamefont {Scardicchio}},\ }\bibfield  {title} {\bibinfo {title} {{Forward approximation as a mean-field approximation for the Anderson and many-body localization transitions}},\ }\href {https://doi.org/10.1103/physrevb.93.054201} {\bibfield  {journal} {\bibinfo  {journal} {Phys. Rev. B}\ }\textbf {\bibinfo {volume} {93}},\ \bibinfo {pages} {054201} (\bibinfo {year} {2016})}\BibitemShut {NoStop}%
\bibitem [{\citenamefont {Scardicchio}\ and\ \citenamefont {Thiery}()}]{Scardicchio2017}%
  \BibitemOpen
  \bibfield  {author} {\bibinfo {author} {\bibfnamefont {A.}~\bibnamefont {Scardicchio}}\ and\ \bibinfo {author} {\bibfnamefont {T.}~\bibnamefont {Thiery}},\ }\bibfield  {title} {\bibinfo {title} {{Perturbation theory approaches to Anderson and Many-Body Localization: some lecture notes}},\ }\href {https://doi.org/10.48550/arXiv.1710.01234} {\bibinfo  {journal} {arXiv:1710.01234}\ }\BibitemShut {NoStop}%
\bibitem [{\citenamefont {De~Tomasi}\ \emph {et~al.}(2017)\citenamefont {De~Tomasi}, \citenamefont {Bera}, \citenamefont {Bardarson},\ and\ \citenamefont {Pollmann}}]{deTomasiQMI}%
  \BibitemOpen
\bibfield  {journal} {  }\bibfield  {author} {\bibinfo {author} {\bibfnamefont {G.}~\bibnamefont {De~Tomasi}}, \bibinfo {author} {\bibfnamefont {S.}~\bibnamefont {Bera}}, \bibinfo {author} {\bibfnamefont {J.~H.}\ \bibnamefont {Bardarson}},\ and\ \bibinfo {author} {\bibfnamefont {F.}~\bibnamefont {Pollmann}},\ }\bibfield  {title} {\bibinfo {title} {Quantum mutual information as a probe for many-body localization},\ }\href {https://doi.org/10.1103/PhysRevLett.118.016804} {\bibfield  {journal} {\bibinfo  {journal} {Phys. Rev. Lett.}\ }\textbf {\bibinfo {volume} {118}},\ \bibinfo {pages} {016804} (\bibinfo {year} {2017})}\BibitemShut {NoStop}%
\bibitem [{\citenamefont {Villalonga}\ and\ \citenamefont {Clark}(2020{\natexlab{a}})}]{villalonga2020characterizing}%
  \BibitemOpen
  \bibfield  {author} {\bibinfo {author} {\bibfnamefont {B.}~\bibnamefont {Villalonga}}\ and\ \bibinfo {author} {\bibfnamefont {B.~K.}\ \bibnamefont {Clark}},\ }\href {https://arxiv.org/abs/2007.06586} {\bibinfo {title} {Characterizing the many-body localization transition through correlations}} (\bibinfo {year} {2020}{\natexlab{a}}),\ \Eprint {https://arxiv.org/abs/2007.06586} {arXiv:2007.06586 [cond-mat.dis-nn]} \BibitemShut {NoStop}%
\bibitem [{\citenamefont {Morningstar}\ \emph {et~al.}(2022)\citenamefont {Morningstar}, \citenamefont {Colmenarez}, \citenamefont {Khemani}, \citenamefont {Luitz},\ and\ \citenamefont {Huse}}]{Morningstar2022}%
  \BibitemOpen
  \bibfield  {author} {\bibinfo {author} {\bibfnamefont {A.}~\bibnamefont {Morningstar}}, \bibinfo {author} {\bibfnamefont {L.}~\bibnamefont {Colmenarez}}, \bibinfo {author} {\bibfnamefont {V.}~\bibnamefont {Khemani}}, \bibinfo {author} {\bibfnamefont {D.~J.}\ \bibnamefont {Luitz}},\ and\ \bibinfo {author} {\bibfnamefont {D.~A.}\ \bibnamefont {Huse}},\ }\bibfield  {title} {\bibinfo {title} {Avalanches and many-body resonances in many-body localized systems},\ }\href {https://doi.org/10.1103/physrevb.105.174205} {\bibfield  {journal} {\bibinfo  {journal} {Phys. Rev. B}\ }\textbf {\bibinfo {volume} {105}},\ \bibinfo {pages} {174205} (\bibinfo {year} {2022})}\BibitemShut {NoStop}%
\bibitem [{\citenamefont {Khemani}\ \emph {et~al.}(2017)\citenamefont {Khemani}, \citenamefont {Lim}, \citenamefont {Sheng},\ and\ \citenamefont {Huse}}]{Khemani2017MBLT}%
  \BibitemOpen
  \bibfield  {author} {\bibinfo {author} {\bibfnamefont {V.}~\bibnamefont {Khemani}}, \bibinfo {author} {\bibfnamefont {S.~P.}\ \bibnamefont {Lim}}, \bibinfo {author} {\bibfnamefont {D.~N.}\ \bibnamefont {Sheng}},\ and\ \bibinfo {author} {\bibfnamefont {D.~A.}\ \bibnamefont {Huse}},\ }\bibfield  {title} {\bibinfo {title} {Critical properties of the many-body localization transition},\ }\href {https://doi.org/10.1103/PhysRevX.7.021013} {\bibfield  {journal} {\bibinfo  {journal} {Phys. Rev. X}\ }\textbf {\bibinfo {volume} {7}},\ \bibinfo {pages} {021013} (\bibinfo {year} {2017})}\BibitemShut {NoStop}%
\bibitem [{\citenamefont {Villalonga}\ and\ \citenamefont {Clark}(2020{\natexlab{b}})}]{Villalonga2020}%
  \BibitemOpen
  \bibfield  {author} {\bibinfo {author} {\bibfnamefont {B.}~\bibnamefont {Villalonga}}\ and\ \bibinfo {author} {\bibfnamefont {B.~K.}\ \bibnamefont {Clark}},\ }\bibfield  {title} {\bibinfo {title} {Eigenstates hybridize on all length scales at the many-body localization transition},\ }\href@noop {} {\bibfield  {journal} {\bibinfo  {journal} {arXiv:2005.13558}\ } (\bibinfo {year} {2020}{\natexlab{b}})}\BibitemShut {NoStop}%
\bibitem [{\citenamefont {Ha}\ \emph {et~al.}(2023)\citenamefont {Ha}, \citenamefont {Morningstar},\ and\ \citenamefont {Huse}}]{Ha2023}%
  \BibitemOpen
  \bibfield  {author} {\bibinfo {author} {\bibfnamefont {H.}~\bibnamefont {Ha}}, \bibinfo {author} {\bibfnamefont {A.}~\bibnamefont {Morningstar}},\ and\ \bibinfo {author} {\bibfnamefont {D.~A.}\ \bibnamefont {Huse}},\ }\bibfield  {title} {\bibinfo {title} {{Many-Body Resonances in the Avalanche Instability of Many-Body Localization}},\ }\href {https://doi.org/10.1103/physrevlett.130.250405} {\bibfield  {journal} {\bibinfo  {journal} {Phys. Rev. Lett.}\ }\textbf {\bibinfo {volume} {130}},\ \bibinfo {pages} {250405} (\bibinfo {year} {2023})}\BibitemShut {NoStop}%
\end{thebibliography}%
\end{document}